%% file: wileyNJD-AMA.tex
\documentclass[AMA,STIX1COL]{WileyNJD-v2}
\usepackage{longtable}
\usepackage{booktabs,graphicx,pdflscape}
\renewcommand{\printjnlcitation}{}
\historydates{}
\doiheadtext{}
\doi{}

\articletype{Research Article}%

\begin{document}

\title{A model-assisted design for phase I oncology drug combination trials incorporating backfill}

\author[1]{Connor Celum*}
\author[2,3]{Evan M. Bagley}

\authormark{CELUM AND BAGLEY}

\address[1]{%
  \orgdiv{Statistics Department},
  \orgname{California Polytechnic State University},
  \orgaddress{San Luis Obispo, CA, \country{USA}}
}

\address[2]{%
  \orgdiv{Department of Public Health Sciences},
  \orgname{Medical University of South Carolina},
  \orgaddress{Charleston, SC, \country{USA}}
}

\address[3]{%
  \orgdiv{Hollings Cancer Center},
  \orgname{Medical University of South Carolina},
  \orgaddress{Charleston, SC, \country{USA}}
}

\corres{%
  *Connor Celum, Statistics Department,
  Faculty Offices East, Building 25,
  San Luis Obispo, CA 93405.
  \email{ccelum@calpoly.edu}
}

\abstract[Abstract]{Historically, phase I oncology trials were designed with the focus to find the highest tolerable dose. However, given the FDA's acknowledgment that the highest tolerable dose might not be the dose which is optimal biologically, there is an increased emphasis the development of designs that explore the dose-toxicity and dose-efficacy relationship. Backfill designs have emerged as a popular response to address this need by allocating participants to doses deemed safe while awaiting participants outcomes. Previous designs have explored how to assign participants to doses in backfill designs in a rigorous manner while providing simplicity and performance (a class of designs known as model-assisted) for single agents. Still, a research gap exists in the lack of model-assisted backfill designs for drug combinations. Frequently, the combinations two of cancer agents are explored in phase I trials, with the goal of identifying the highest tolerable combination. This paper incorporates and extends methodology based on previous model-assisted backfill designs to provide clinicians with a simple-to-use backfill design for phase I drug combination trials.}


\fundinginfoheadtext{}
\fundingInfo{%
  This work was supported in part by 1)
  Medical University of South Carolina Hollings Cancer
  Center Data Science Shared Resource (P30 CA138313)
}

\maketitle

\footnotetext{\textbf{Abbreviations:} DLT, dose-limiting toxicity; MTD, maximum tolerated dose; FDA, Food and Drug Administration; BOIN, Bayesian optimal interval; BF-BOIN, backfill Bayesian optimal interval;
BOIN-Comb, Bayesian optimal interval for drug combinations;
BF-BOIN-Comb, backfill Bayesian optimal interval for drug combinations}

\section{Introduction}
\label{sec:introduction}

Phase I cancer trials are safety studies testing dose levels to identify the maximum tolerated dose (MTD), defined as the dose level with a dose-limiting toxicity (DLT) rate closest to a prespecified target. These trials are adaptive in that participants are assigned to dose levels based on previous participants’ DLT outcomes. Well-designed studies that correctly select the true MTD are a critical aspect of the drug development pipeline.\cite{conaway2019impact} Frequently, multiple drugs are of interest, and different combinations of dose levels are evaluated during dose escalation. A recent review\cite{chihara2022early} showed that 69\% of National Cancer Institute sponsored phase I trials for solid tumors used some form of drug combination therapy. In an era of targeted therapies, this trend will likely continue, motivating the need for drug combination designs tailored to the study’s objectives \cite{jin2023rational}.

Additionally, backfill is a design component that is growing in popularity and is defined as enrolling participants at previously evaluated dose levels that have been cleared for safety and have shown preliminary evidence of activity.\cite{okusanya2025fda} Backfill is an important consideration in the context of the FDA’s Project Optimus, as the highest tolerable dose is not always the same as the optimal dose to move forward in drug development.\cite{qiu2025robust, FDA2024DoseOptimization} However, backfill is often ill-defined in protocols, motivating the need for innovative dose-finding designs that incorporate backfill responses. The proposed design will accommodate backfill responses, in the context for drug combinations.

Previous model-based and model-assisted dose-finding designs in drug combination studies include Wages et al. \cite{wages2011continual}, Lin \& Yin \cite{lin2017bayesian}, and Bagley \& Wages  \cite{bagley2025dose}. In recent years, several backfill dose-finding designs have been proposed.\cite{dehbi2021controlled,zhao2024backfilling,chen2026boin,barnett2023backfilling} However, these backfill designs were developed for the single-agent setting. This article proposes the BF-BOIN-Comb design, a phase I dose-escalation design for drug combinations that allows for backfilling of participants at lower dose combinations.

\section{Methodology}
\label{sec:methodology}

The proposed method extends the Backfill Bayesian Optimal Interval Design (BF-BOIN) of Zhao et al.\cite{zhao2024backfilling}, while using the framework of the Bayesian Optimal Interval Design for Dose Combinations (BOIN-Comb) from Lin \& Yin\cite{lin2017bayesian}, to accommodate multi-drug dose escalation. We first describe the BOIN framework for single-agent dose escalation from Liu \& Yuan \cite{liu2015bayesian}, then review its extension to incorporate backfill, and finally introduce the proposed method, BF-BOIN-Comb.

\subsection{Bayesian optimal interval method (BOIN)}
\label{sec:boin}

Let \(\theta^{*}\) denote the target DLT rate, typically between 0.20 and 0.33, and let \(\lambda_e\) and \(\lambda_d\) denote the escalation and de-escalation boundaries, respectively. Dose-assignment decisions are made by comparing the observed DLT rate at the current dose level with these boundaries. If the observed DLT rate at the current dose level, \(\hat{p}_{c}\), is not more than \(\lambda_e\), escalate one dose level; if the DLT rate is contained in the interval \((\lambda_e, \lambda_d)\), retain the current dose level; otherwise, the DLT rate exceeds \(\lambda_d\) and the dose level is de-escalated one dose level.

\begin{itemize}
  \item If \(\hat{p}_c \leq \lambda_e\), escalate one dose level
  \item If \(\lambda_e \leq \hat{p}_{c} \leq \lambda_d\),
        retain the current dose level
  \item If \(\hat{p}_c \geq \lambda_d\),
        de-escalate one dose level
\end{itemize}

Details regarding the specification and derivation of these boundaries are provided in Liu and Yuan.\cite{liu2015bayesian} The design enrolls participants until the maximum number of participants is enrolled, denoted as \(N_{\mathrm{max}}\), is reached. At the conclusion of the study, DLT rates across dose levels are estimated using isotonic regression via the pooled adjacent violators algorithm \cite{barlow1972isotonic}, ensuring that the estimated rates are nondecreasing with respect to the increasing dose levels.

\subsection{Bayesian optimal interval method extended for backfill (BF-BOIN)}
\label{sec:bf-boin}

Zhao et al.\cite{zhao2024backfilling} introduced a backfill extension of BOIN for single-agent dose-escalation studies. The design retains the original BOIN decision rules for dose escalation while allowing additional participants to enroll at doses open for backfill. Two key components of the design are (1) determining when to open and close doses for backfill and (2) addressing observed DLT rates at lower doses which lead to conflicting decisions, relative to the escalation decision at the current dose. The second component (2) is of particular importance in a model-assisted setting where dose escalation decisions are made locally at the current dose level.

Under this method, dose level \(b\) is eligible for backfill if \(b<c\) (i.e., the dose has been cleared for safety), where \(c\) denotes the current dose level, and at least one response has been observed at or below dose level \(b\) (i.e., some signal of efficacy is observed). The response criterion is specific to the protocol and disease setting but should demonstrate a preliminary signal of clinical activity to support backfill at doses that may offer therapeutic benefit. For our purposes, response is treated as a binary outcome. For example, in solid tumor studies, objective response may be defined as an observed radiographic complete or partial response.

To account for additional DLTs observed during backfill, BF-BOIN closes dose level \(b\) to further backfill enrollment if both (1) the observed DLT rate at that dose, calculated using all participants enrolled in the escalation and backfill cohorts, exceeds the de-escalation boundary \(\lambda_d\) and (2) the pooled DLT rate across dose levels \(b\) and \(b+1\) also exceeds this boundary. The second criteria for closing a dose for backfill is added to account for a possible false signal of toxicity, which can result from the simultaneous testing of several doses below the current dose. By requiring overly toxic signals to come from both (1) dose level \(b\), and (2) from the pooled DLT rate at dose level \(b\) and one dose level higher, BF-BOIN aims to control the rate of false toxicity signals.

The BF-BOIN design provides a mechanism to address occurrences of accumulated backfill responses leading to conflicting decisions, relative to the current dose. Such a situation would arise if participant data accumulated at the current dose suggests dose escalation while backfill responses at a lower dose contributed to the suggestion of dose de-escalation. The table below, as adapted from Qiu \& Li\cite{qiu2025robust}, highlights all such conflicts. When such a conflict arises, pooled averages are used to reconcile such conflicts in a data-driven manner, while also preserving the model-assisted nature of BOIN.

\begin{table}[htbp]
  \centering
  \setlength{\fboxsep}{3pt}
  \setlength{\fboxrule}{0.5pt}
  \fbox{%
    \includegraphics[width=0.75\linewidth]{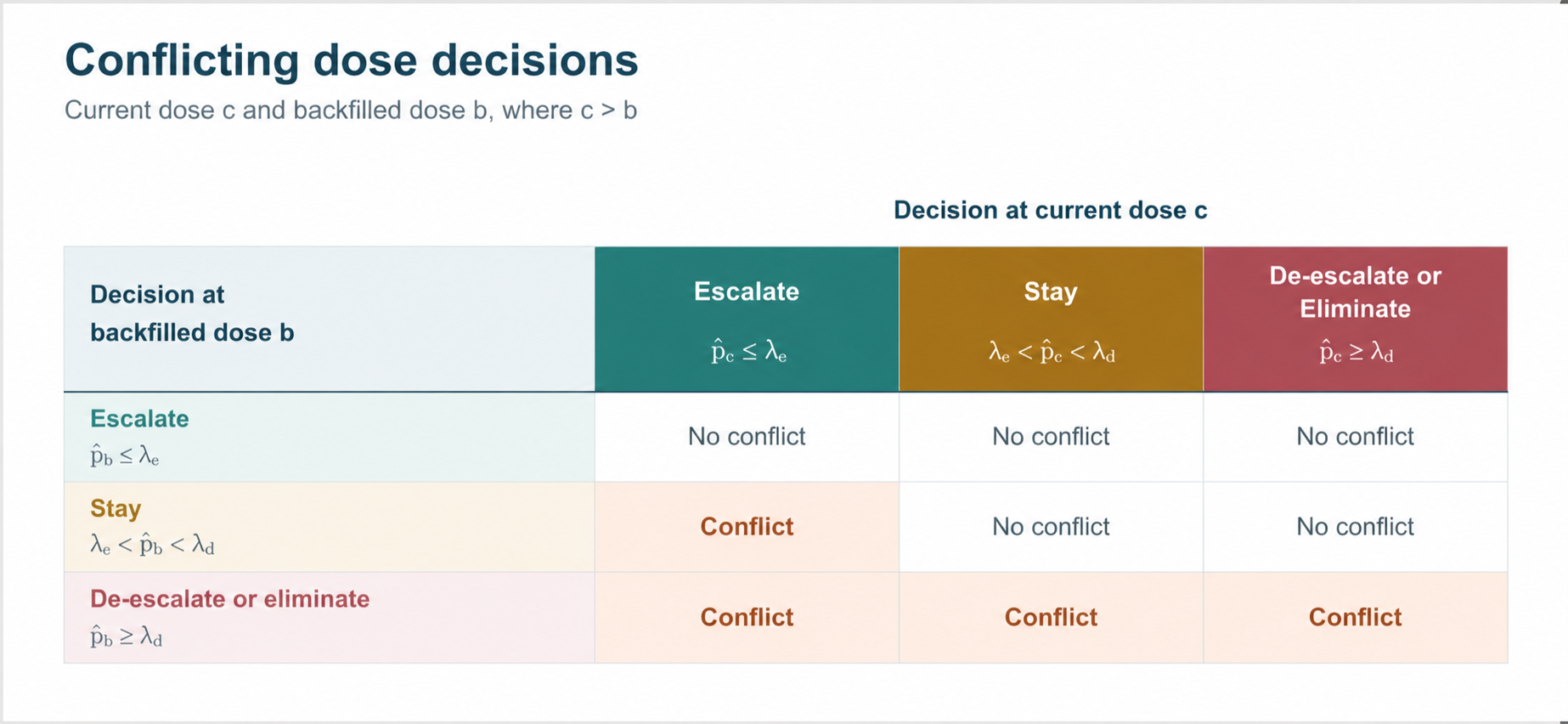}%
  }
  \caption{Conflicting dose-assignment decisions. The following image was generated with the assistance from OpenAI's image-generation tool, accessed through Codex on October 9, 2026. This content was review and corrected by the authors.}
  \label{tab:conflict}
\end{table}
Together, these extensions incorporate backfill enrollment into the BOIN framework while allowing accumulating DLT and response data to inform backfill decisions. The operating characteristics of BF-BOIN were evaluated through simulation to assess its performance across a range of scenarios.

\subsection{Proposed method.}
\label{sec:proposed-method}

The proposed method extends the BF-BOIN framework to the drug combination setting by incorporating methodology used in BOIN for drug combinations.\cite{lin2017bayesian} The proposed method retains BOIN decision boundaries while accounting for the partial ordering of the drug combination DLT probabilities. Backfill enrollment is permitted at dose combinations that are less than the current dose-combination. Accumulating data from both backfill and escalation cohorts contribute to subsequent dose assignments.

Consider agents A and B, with dose levels \(i=1,\ldots,I\) for drug A and \(j=1,\ldots,J\) for drug B. Further, let \(d_{ij}\) denote the combination of the \(i\)th dose of agent A and the \(j\)th dose of agent B. Let \(p_{ij}\) be the DLT probability corresponding to \(d_{ij}\) and \(\theta^{*}\) be the target DLT probability. We assume that toxicity is nondecreasing as the dose of either agent increases while the other is held fixed. Thus, \(p_{i'j'} \leq p_{ij}\) when \(i' \leq i\) and \(j' \leq j\). Prior to the trial initiation, combinations involving a higher dose of one agent and a lower dose of the other agent (i.e. \(d_{12}\) and \(d_{21}\)) do not have a known DLT probability ordering.

Following the BOIN design, participants are enrolled in cohorts (often of size three), and the trial is initiated at the lowest dose combination \(d_{11}\). Dose escalation occurs after all participants have completed the DLT assessment, meaning each participant has either been observed for the entire assessment window (often one treatment cycle or 28-days) and did not experience a DLT, or experienced a DLT within the assessment window. Let \(n_{ij}\) and \(y_{ij}\) denote the number of participants with complete DLT assessments and number of DLT outcomes at dose combination \(d_{ij}\). Letting \(c = d_{ij}\) denote the current dose combination, if \(\hat{p}_{ij} =\frac{y_{ij}}{n_{ij}} \le \lambda_e\) dose escalation occurs, if \(\hat{p}_{ij} \ge \lambda_{d}\) dose de-escalation occurs, and if \(\hat{p}_{ij} \in (\lambda_e, \lambda_d)\) the dose combination remains the same for the subsequent cohort.
\begin{itemize}
    \item If the decision is dose escalation, the escalation candidates are \(d_{i+1,j}\) and \(d_{i,j+1}\).
    \item If the decision is dose de-escalation, the de-escalation candidates are \(d_{i-1,j}\) and \(d_{i,j-1}\)
\end{itemize}
When two candidates are available, the dose combination with the higher posterior probability of lying within the interval \((\lambda_e,\lambda_d)\) is selected. Following Lin and Yin,\cite{lin2017bayesian} the \(\operatorname{Beta}(0.5,0.5)\) prior is used, giving
\begin{equation*}
p_{ij} \mid y_{ij},n_{ij}
\sim
\operatorname{Beta}
\bigl(y_{ij}+0.5,\;n_{ij}-y_{ij}+0.5\bigr),
\label{eq:posterior}
\end{equation*}
and the selection score is defined as:
\begin{equation}
s_{ij}
=
\Pr\bigl(\lambda_e < p_{ij} < \lambda_d|y_{ij},n_{ij}\bigr).
\label{eq:selection-score}
\end{equation}
Ties are resolved by equal-probability randomization among the maximizing candidates. Selection scores use all complete DLT assessments at each candidate which may include observations from backfill participants. If no movement is possible at a grid boundary, i.e. de-escalate from \(d_{11}\) or escalate from \(d_{IJ}\), the current combination is retained, subject to safety stopping. Note, all possible selection scores can be outlined in a table prior to trial initiation, as can be seen in corresponding shiny app for BOIN-Comb. \cite{Zhou2026}

\subsubsection{Backfill eligibility and allocation}
\label{sec:backfill-eligibility}

Enrollment of the escalation cohort precedes allocating participants to backfill doses. Participants are assigned to backfill doses after the escalation cohort is fully enrolled and while DLT assessments are awaiting completion. Let \(d_{ij}\) denote the current dose combination \(c\). For a dose combination, \(b=d_{i'j'}\), to be open for backfill:

\begin{enumerate}
  \item The dose combination must be lower in at least one agent and cannot be higher in either agent. That is, \(i'\leq i\) and \(j'\leq j\), with at least one of \(i'<i\) or \(j'<j\).

  \item There has been at least one response at \(d_{i'j'}\) or a lower dose combination, \(d_{i^{*}j^{*}}\), where \(i^{*}\leq i\) and \(j^{*}\leq j\). 
\end{enumerate}

DLTs accumulating during backfill may indicate that dose combination \(b=d_{i'j'}\) is unsafe for backfill. Dose combination \(b=d_{i'j'}\) is declared unsafe for backfill if both of the following conditions are met:

\begin{enumerate}
  \item The DLT rate at \(b=d_{i'j'}\) demonstrates excessive toxicity:
  \begin{equation}
    \frac{y_{i'j'}}{n_{i'j'}} \geq \lambda_d
    \label{eq:backfill-local-toxicity}
  \end{equation}

  \item The pooled DLT rate at \(b=d_{i'j'}\) and dose-combinations immediately higher demonstrates excessive toxicity.
\end{enumerate}

To more precisely describe the second of these conditions, define:
\begin{equation*}
\mathcal{L}_{b} = \{d_{i'j'}\} \cup \{d_{i'+1,j'}:i'<i\} \cup \{d_{i',j'+1}:j'<j\} 
\end{equation*}

Letting \(y_b^{+}=\sum_{d\in \mathcal{L}_b}y_d\) and
\(n_b^{+}=\sum_{d\in \mathcal{L}_b}n_d\), the second condition is met when:
\begin{equation}
  \frac{y_b^{+}}{n_b^{+}} \geq \lambda_d
  \label{eq:backfill-pooled-toxicity}
\end{equation}

If dose combination \(b=d_{i'j'}\) is closed to backfill due to accumulating DLTs demonstrating that the combination is overly toxic, based on the two criteria above, all higher dose combinations, \(d_{i^{*}j^{*}}\) where \(i'\leq i^{*}\) \& \(j'\leq j^{*}\), are closed to backfill.

Among eligible combinations, the allocation rule selects uniformly from the maximal dose combinations. A dose combination is maximal if no other eligible combination is at least as high in both agents and strictly higher in one. An alternative allocation rule is randomizing uniformly among all eligible combinations.

Similar to BF-BOIN, a dose combination is closed to backfill once the total participants assigned to that combination exceeds \(n_{\mathrm{cap}}\). We define \(N_{\mathrm{esc}}\) to be the maximum number of escalation cohorts. If the dose-escalation decision is ``stay'' at the current dose combination and the total participants observed at this combination exceeds \(n_{\mathrm{stop}}\), the dose escalation phase ends.

\subsubsection{Dose Escalation and Backfill DLT Outcomes}
\label{sec:conflict-resolution}

\begin{figure}[htbp]
  \centering
  \setlength{\fboxsep}{3pt}
  \setlength{\fboxrule}{0.5pt}
  \fbox{%
    \includegraphics[
      width=\dimexpr\linewidth-2\fboxsep-2\fboxrule\relax
    ]{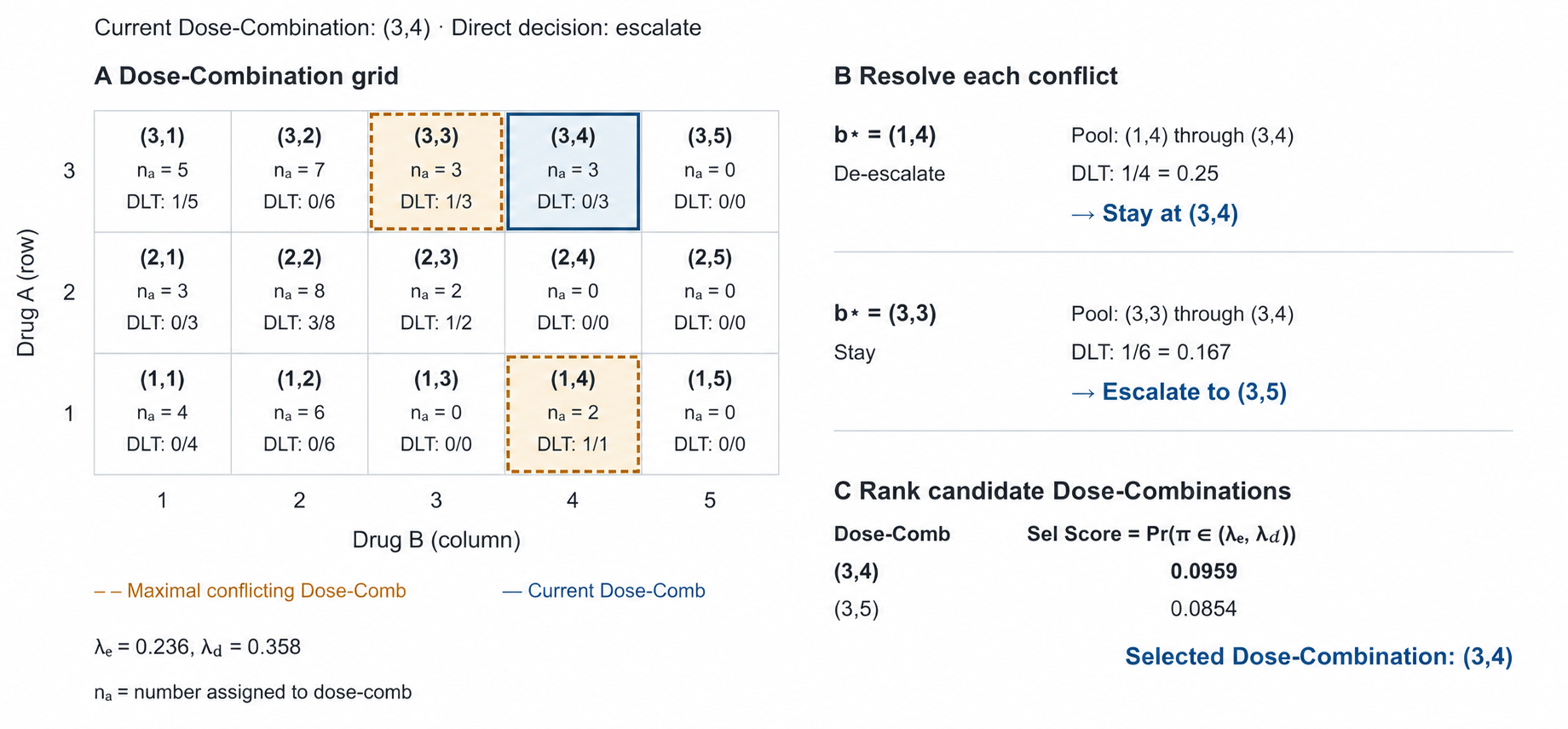}%
  }
  \caption{Illustration of conflict resolution at dose-combination $d_{(3,4)}$. The following image was generated with the assistance from OpenAI's image-generation tool, accessed through Codex on October 9, 2026. This content was review and corrected by the authors.}
  \label{fig:conflict-resolution}
\end{figure}

Accumulating DLTs occurring during backfill better inform the toxicity profile of dose combinations below the current dose combination. Like previous model-assisted backfill designs,\cite{zhao2024backfilling,chen2026boin} the challenge arising from accumulating backfill data leading to a conflicting decision, as outlined in Table~\ref{tab:conflict}, needs to be addressed. For example (see Figure \ref{fig:conflict-resolution}), this can arise if the current dose combination is \(d_{34}\) and all participants have completed the DLT assessment which leads to the decision to escalate, while at the backfill dose combination  \(d_{33}\) the accumulated DLTs suggest de-escalation. This section will describe how the BF-BOIN-Comb design utilizes the original BF-BOIN framework to address such conflicts.

Suppose there exists at least one backfill dose combination \(b\) below the current dose combination \(c=d_{ij}\), leading to a conflicting decision. Let \(B^{\star}\) denote the set of maximal backfill dose combinations that lead to conflicting decisions. For an illustrative example, see Figure~\ref{fig:conflict-resolution}. In this example, the current dose combination is \(d_{34}\) and after all participants from the most recent escalation cohort have completed DLT assessment, the decision is to escalate. However, there are conflicting decisions coming from dose combinations \(\{d_{14},d_{22},d_{23},d_{33}\}\), corresponding to decisions \(\{\text{De-escalate},\text{De-escalate},\text{De-escalate},\text{Stay}\}\), respectively. Recalling the definition of maximal dose-combination in Section~\ref{sec:backfill-eligibility}, the set of maximal conflicting dose-combinations is \(B^{\star}=\{d_{33},d_{14}\}\).

After obtaining \(B^{\star}\), for each \(b^{\star}=d_{i^{\star}j^{\star}}\in B^{\star}\), we define a pooled DLT rate of all dose-combinations in the rectangle between \(b^{\star}\) and \(c\). First, define \(R_{(b^{\star},c)}\) as the set of all such dose-combinations:
\begin{equation*}
R_{(b^{\star},c)}
= 
\bigl\{d_{i'j'}:
i^{\star}\leq i'\leq i
\;\&\;
j^{\star}\leq j'\leq j,
\text{ where } b^{\star}=d_{i^{\star}j^{\star}}
\text{ and } c=d_{ij}
\bigr\}.
\label{eq:conflict-pooling-region}
\end{equation*}

For example, \(R_{(d_{33},c=d_{34})}=\{d_{33},d_{34}\}\) and \(R_{(d_{14},c=d_{34})}=\{d_{14},d_{24},d_{34}\}\). 
Then, define
\begin{equation}
\hat{q}_{(b^{\star},c)}
=
\frac{\displaystyle\sum_{d\in R_{(b^{\star},c)}} y_d}
     {\displaystyle\sum_{d\in R_{(b^{\star},c)}} n_d}.
\label{eq:conflict-pooled-rate}
\end{equation}

For example, \(\hat{q}_{(d_{14},c)}=0.25\) and \(\hat{q}_{(d_{33},c)}=0.167\). Based on \(\hat{q}_{(b^{\star},c)}\) an admissible candidate dose-combination is added to the set of candidate dose combinations, \(\mathbf{C}_d\).

\begin{itemize}
  \item If \(\hat{q}_{(b^{\star},c)}\leq\lambda_e\), add the escalation recommendation (\(d_{i+1,j}\) or \(d_{i,j+1}\), which ever has a higher selection score) to \(\mathbf{C}_d\).

  \item If \(\hat{q}_{(b^{\star},c)}\geq\lambda_d\), add maximal dose combination(s) \(d_{i^{\prime}j^{\prime}}\in R_{(b^{\star},c)}\), with \(\hat{q}_{(b^{\star},d_{i^{\prime}j^{\prime}})}\leq\lambda_d\), if such a dose combination does not exist, add the de-escalation recommendation \(d_{i^{\star}-1,j^{\star}}\) or \(d_{i^{\star},j^{\star}-1}\), whichever has a higher selection score, to \(\mathbf{C}_d\).

  \item Else, \(\hat{q}_{(b^{\star},c)}\in(\lambda_e,\lambda_d)\) and the current dose combination is added to \(\mathbf{C}_d\)
\end{itemize}

As \(\hat{q}_{(d_{14},c)}=0.25\in(0.236,0.358)\), the decision to stay at \(d_{34}\) is added to \(\mathbf{C}_d\). Additionally, as \(\hat{q}_{(d_{33},c)}=0.167\leq0.236\), the decision to escalate to \(d_{35}\) is added to \(\mathbf{C}_d\), resulting in \(\mathbf{C}_d=\{d_{34},d_{35}\}\).

The subsequent cohort will be assigned to the dose-combination in \(\mathbf{C}_d\) with the highest selection score. In our running example, \(d_{34}\) has the highest selection score, and the subsequent cohort will be observed at this combination.
\subsubsection{Final Dose selection}
\label{sec:final-dose-selection}

The proposed design uses the BOIN elimination rule to eliminate overly toxic dose combinations, namely combinations with a posterior probability of at least 95\% of excessive toxicity. A combination meeting this criterion is eliminated from the trial, along with all higher combinations. Elimination of the lowest combination terminates the trial and no MTD is selected.

Dose escalation ends when the prespecified maximum number of escalation cohorts is reached (\(N_{\mathrm{esc}}\)) or dose escalation decision is to remain at the current combination and at least \(n_{\mathrm{stop}}\) participants have been assigned. Backfill accrual ends with escalation. All enrolled backfill participants are followed until DLT outcomes are resolved. MTD selection uses both escalation and backfill DLT via bivariate isotonic regression,\cite{dykstra1982algorithm} enforcing nondecreasing toxicity estimates along both dose dimensions. The procedure selects a single MTD combination. Efficacy and toxicity data governs backfill eligibility, while final MTD selection is based on toxicity alone.

\section{Results}
\label{sec:results}
Comparisons between BF-BOIN-Comb and BOIN-Comb were obtained through simulation studies. The operating characteristics of the BF-BOIN-Comb design were evaluated in 16 paired toxicity and response scenarios. To avoid cherry picking scenarios partial to the proposed design, scenarios were selected from the previous BOIN-Comb study \cite{lin2017bayesian} and from Wages and Conaway.\cite{wages2014phase}. Scenarios 1 through 10 use the toxicity scenarios of the BOIN-Comb study \cite{lin2017bayesian}, formed by three dose levels of agent $A$ and five dose levels of agent $B$. Since Lin \& Yin \cite{lin2017bayesian} only included toxicity scenarios, response profiles were obtained from the first 10 scenarios in Zhao et al. \cite{zhao2024backfilling}, assigning these corresponding efficacy profiles to dose combinations at dose level one in agent $A$ and adding 0.05 for each increase in dose level in agent $A$. Scenarios 11 to 16 use the toxicity and response profiles in Wages and Conaway \cite{wages2014phase} with three dose levels of agent $A$ and three dose levels of agent $B$. The toxicity/response profiles are provided in Table~\ref{tab:simulation-scenarios}.

%Scenarios 1 through 10 comprise 15 dose combinations formed by five dose levels of one agent and three of the other, 

%whereas Scenarios 11--16 comprised nine combinations formed by a three by three grid of dose levels for each agent. The scenarios varied in the location and number of combinations closest to the target DLT probability and included increasing and plateauing response patterns. Scenario 11 represented a setting in which all combinations had DLT probabilities below the target, whereas Scenario 16 represented a setting in which all combinations were overly toxic. The scenario-specific toxicity and response probabilities are presented in [Supplemental Table XXX].

The target DLT probability was set to \(\theta^{*}=0.30\). Participants were enrolled in cohorts of size three, with maximum escalation sample sizes of 30, 45, or 60 participants, corresponding to 10, 15, or 20 cohorts, respectively. For each combination of scenario and maximum escalation sample size, 5,000 trials were simulated for each design. BOIN boundaries were specified using lower and upper reference toxicity probabilities of 0.18 and 0.42, yielding escalation and de-escalation boundaries of approximately 0.236 and 0.358. These values correspond to recommendations given in the original BOIN paper.\cite{yuan2016bayesian} 
Escalation also halted when the next recommended combination was unchanged and at least 9 participants had been assigned to that combination across escalation and backfill (i.e. \(n_{\mathrm{stop}}=9\)). After 12 participants have been observed at a dose combination, the dose combination was closed to backfill (i.e. \(n_{\mathrm{cap}}=12\)).

When comparing BF-BOIN-Comb and BOIN-Comb, the same simulation framework, escalation and stopping rules, and final MTD-selection procedure were utilized. In scenarios 1--15, correct selection was defined as selecting any combination whose true DLT probability was closest to 0.30, allowing multiple correct combinations when these distances were tied. Across these scenarios, the following operating characteristics were recorded: (1) percent correct selection, (2) accuracy index (AI) \cite{cheung2011dose}, (3) over-dosing selection, (4) number and percent of participants assigned dose combinations below the MTD, (5) number and percent of participants assigned to the MTD or MTDs, (6) number and percent of participants assigned dose combinations above the MTD, (7) average sample size, and (8) average duration in months. These metrics are selected to match the selection of operating characteristics in Zhao et al \cite{zhao2024backfilling}, while adding AI to allow for a complete comparison of dose-selection across all dose combinations. Note that AI only includes simulations that resulted in the selection of a dose combination and excludes trials terminated early due to excessive toxicity. Scenario 16 presents a scenario where all dose combinations are excessively toxic and the correct decision is to select no MTD. As a result, we consider this scenario separately, considering percentage of simulations where no MTD is selected, mean sample size, and mean trial duration. Appendix subsection \ref{app:sim_result} provides these results with Table \ref{tab:oc-30}, Table \ref{tab:oc-45}, and Table \ref{tab:oc-60}, providing the operating characteristics for scenarios 1--15 with the maximum enrollment cohort set to 30, 45, and 60, respectively. Additionally, Table \ref{tab:scenario16-nstop9} provides the results from scenario 16.

In comparing dose combination selection performance using percent correct selection and AI across the two methods: 
\begin{enumerate}
    \item[$\bullet$] BF-BOIN-Comb consistently outperforms BOIN-Comb in scenarios: 1, 2, 4, 7, 8, 9 \& 14
    \item[$\bullet$] BOIN-Comb consistently outperforms BF-BOIN-Comb in scenarios: 5, 6, 11, 12, \& 13
    \item[$\bullet$] Mixed results across $N_{esc}$, AI, and percent correct selection in scenarios: 3, 10, \& 15
\end{enumerate}
Additionally, the statistics of overdosing selection percentage and number/percentage of participants above the MTD demonstrate that the BF-BOIN-Comb tends to increase participants safety. The principal benefit of the BF-BOIN-Comb design is to increase the number of participants observed at dose combinations at or below the MTD, aiding in selection of the biologically optimal dose. Zhao et al. \cite{zhao2024backfilling} provided similar dose selection results with BF-BOIN both increasing and decreasing percent correct selection relative to the original BOIN design, depending on the scenario.

%We also summarized participant allocation across the dose grid, mean total sample size, and mean trial duration, measured from the first enrollment until the final DLT outcome was resolved. For the all-toxic Scenario 16, particular attention was given to the probabilities of early termination for toxicity and of recommending no MTD.
\subsection{Sensitivity Analysis}\label{sub:sen_ana}
% Proofread

In the proposed BF-BOIN-Comb design, adjustments are made when backfilling (lower dose combinations) leads to a conflicting decision, relative to the escalation decision at the current dose combination. However, it is possible for a dose combination below the current dose combination to lead to a conflicting decision from a previous escalation cohort. To illustrate this, suppose that the following escalation/de-escalation path was observed and no efficacious responses were observed, implying backfill did not occur:  
\begin{enumerate}
    \item The first cohort is observed at $d_{13}$ and one out of the three participants have a DLT, resulting in the decision to stay at $d_{13}$.
    \item Another cohort is observed and two out of the three participants have a DLT, resulting in the DLT rate $\hat{p}_{13} = \frac{3}{6} = 0.5$ and the decision to de-escalate to dose-combination $d_{12}$.
    \item A cohort is observed at $d_{12}$ and the cumulative responses at $d_{12}$ result in the decision to escalate dose-combination. Dose combination $d_{22}$ has a higher selection score than dose combination $d_{13}$ and the next cohort is assigned dose combination $d_{22}$.
    \item A cohort is observed at $d_{22}$ and the decision is to escalate to $d_{23}$.
    \item A cohort is observed at $d_{23}$, no DLTs are observed and $\hat{p}_{23} = \frac{0}{3} \le \lambda_e$, and the decision is to escalate..
\end{enumerate}

From this sequence of escalation/de-escalation decisions and observed cohorts, we have a dose de-escalation decision at $d_{13}$ and a dose escalation decision at the current dose-combination of $d_{23}$, as a result of escalation cohorts alone, implying such a situation can arise in BOIN-Comb. Keeping these possibilities in mind, a sensitivity analysis was conducted, compared the proposed version of BF-BOIN-Comb, which only makes adjustments based on backfill responses (more closely combining BF-BOIN and BOIN-Comb) and a more conservative version that makes escalation adjustments whenever a lower dose-combination has a DLT rate leading to a conflicting decision, including conflicts that arise outside of backfill. These results are supplied in Appendix subsection \ref{app:bf-conflict-comparison}, where the conservative adjustment is called "all lower". The more conservative variation performs similarly to the correction mechanism presented in this paper, with the more conservative method making slightly more conservative selection and allocation decisions.

\section{Summary}
\label{sec:results}

This article proposes an extension of BF-BOIN that, combined with the BOIN-Comb framework, creates the BF-BOIN-Comb design by adapting backfill eligibility, participant allocation, and conflict resolution to a partially ordered grid of dose combinations. As a result, backfill can be utilized for combinations of two drugs while retaining the relative simplicity of model-assisted designs which utilize the BOIN decision boundaries. 

The proposed framework aims to estimate the MTD, while collecting additional data at dose combinations below the MTD. Collecting toxicity and efficacy information at dose combinations below the MTD is of particular interest with respect to the FDA's Project Optimus, which recognizes that MTD is not always the optimal dose combination. Through the development of the BF-BOIN-Comb, clinicians have an easy to use model-assisted design for drug combination trials that allows for the exploration of the dose combination-toxicity and dose combination-efficacy relationship through backfill. 

The main limitation of the proposed method is the performance loss (relative to BOIN-Comb) in locating the MTD when the MTD happens to be the highest drug combination, for example $d_{5,3}$ is the MTD in scenario 5 and in this scenario there is a performance decrease. This suggests that BF-BOIN-Comb is receiving false signals of safe dose combinations being overly toxic during backfill. A future area of research is looking at addressing the problem of backfill in drug combinations from a model-based approach, converting local dosing decisions into global decisions, sharing information across the entire grid of dose-combinations and potentially lowering the rate of such false toxicity alarms.

\section*{Acknowledgments}
In the generation of simulations, Codex from Open AI was utilized to aid in code creation. All code was reviewed and corrected by the authors. 
%This is acknowledgment text.\cite{Kenamond2013} Provide text here. This is acknowledgment text. Provide text here. This is acknowledgment text. Provide text here. This is acknowledgment text. Provide text here. This is acknowledgment text. Provide text here. This is acknowledgment text. Provide text here. This is acknowledgment text. Provide text here. This is acknowledgment text. Provide text here. This is acknowledgment text. Provide text here. 

%\subsection*{Author contributions}

%This is an author contribution text. This is an author contribution text. This is an author contribution text. This is an author contribution text. This is an author contribution text. 

\subsection*{Financial disclosure}

None reported.

\subsection*{Conflict of interest}

The authors declare no potential conflict of interests.

\section*{Supporting information}
Upon publication, supporting code for reproducibility will be available as part of the online article.

\nocite{*}% Show all bib entries - both cited and uncited; comment this line to view only cited bib entries;
\bibliography{wileyNJD-AMA}%

\section{Appendix\label{app1}}%
\input{simulation-scenarios}
\clearpage
\input{main_sim_results}
\clearpage
\input{BFBOINComb_compare}

\end{document}

%% file: simulation-scenarios.tex
% Input fragment. Requires \usepackage{longtable} in the main preamble.
\clearpage
\subsection{Simulation scenarios}
\label{app:simulation-scenarios}
\begingroup
\small
\setlength{\tabcolsep}{3pt}
\renewcommand{\arraystretch}{1.1}
\begin{longtable}{@{}p{0.47\linewidth}@{\hspace{0.06\linewidth}}p{0.47\linewidth}@{}}
\caption{Toxicity and efficacy probabilities for the 16 simulation scenarios are provided below. Rows correspond to Agent 2 dose levels and columns to Agent 1 dose levels. Bold toxicity probabilities are given to MTD dose combinations (i.e. those with DLT probabilities closest to the DLT target of 0.3).}
\label{tab:simulation-scenarios}\\
\hline
\hline
\endfirsthead
\multicolumn{2}{l}{\tablename\ \thetable\ (continued)}\\
\hline
\hline
\endhead
\multicolumn{2}{r}{Continued on next page}\\
\endfoot
\endlastfoot
\multicolumn{2}{c}{\textbf{Scenario 1}}\\*[3pt]
\begin{minipage}[t]{\linewidth}\centering
\textbf{Toxicity probability}\par\smallskip
\begin{tabular}{@{}rrrrrr@{}}
 & \multicolumn{5}{c}{Agent 1 dose level} \\
Agent 2 & 1 & 2 & 3 & 4 & 5 \\ \hline
3 & 0.15 & \textbf{0.30} & 0.45 & 0.50 & 0.60 \\
2 & 0.10 & 0.15 & \textbf{0.30} & 0.45 & 0.55 \\
1 & 0.05 & 0.10 & 0.15 & \textbf{0.30} & 0.45 \\
\hline\end{tabular}\end{minipage}
 & \begin{minipage}[t]{\linewidth}\centering
\textbf{Efficacy probability}\par\smallskip
\begin{tabular}{@{}rrrrrr@{}}
 & \multicolumn{5}{c}{Agent 1 dose level} \\
Agent 2 & 1 & 2 & 3 & 4 & 5 \\ \hline
3 & 0.40 & 0.50 & 0.55 & 0.60 & 0.65 \\
2 & 0.35 & 0.45 & 0.50 & 0.55 & 0.60 \\
1 & 0.30 & 0.40 & 0.45 & 0.50 & 0.55 \\
\hline\end{tabular}\end{minipage}
 \\[10pt]
\multicolumn{2}{c}{\textbf{Scenario 2}}\\*[3pt]
\begin{minipage}[t]{\linewidth}\centering
\textbf{Toxicity probability}\par\smallskip
\begin{tabular}{@{}rrrrrr@{}}
 & \multicolumn{5}{c}{Agent 1 dose level} \\
Agent 2 & 1 & 2 & 3 & 4 & 5 \\ \hline
3 & 0.45 & 0.55 & 0.60 & 0.70 & 0.80 \\
2 & \textbf{0.30} & 0.45 & 0.50 & 0.60 & 0.75 \\
1 & 0.15 & \textbf{0.30} & 0.45 & 0.50 & 0.60 \\
\hline\end{tabular}\end{minipage}
 & \begin{minipage}[t]{\linewidth}\centering
\textbf{Efficacy probability}\par\smallskip
\begin{tabular}{@{}rrrrrr@{}}
 & \multicolumn{5}{c}{Agent 1 dose level} \\
Agent 2 & 1 & 2 & 3 & 4 & 5 \\ \hline
3 & 0.30 & 0.40 & 0.50 & 0.60 & 0.70 \\
2 & 0.25 & 0.35 & 0.45 & 0.55 & 0.65 \\
1 & 0.20 & 0.30 & 0.40 & 0.50 & 0.60 \\
\hline\end{tabular}\end{minipage}
 \\[10pt]
\multicolumn{2}{c}{\textbf{Scenario 3}}\\*[3pt]
\begin{minipage}[t]{\linewidth}\centering
\textbf{Toxicity probability}\par\smallskip
\begin{tabular}{@{}rrrrrr@{}}
 & \multicolumn{5}{c}{Agent 1 dose level} \\
Agent 2 & 1 & 2 & 3 & 4 & 5 \\ \hline
3 & 0.10 & 0.15 & \textbf{0.30} & 0.45 & 0.55 \\
2 & 0.07 & 0.10 & 0.15 & \textbf{0.30} & 0.45 \\
1 & 0.02 & 0.07 & 0.10 & 0.15 & \textbf{0.30} \\
\hline\end{tabular}\end{minipage}
 & \begin{minipage}[t]{\linewidth}\centering
\textbf{Efficacy probability}\par\smallskip
\begin{tabular}{@{}rrrrrr@{}}
 & \multicolumn{5}{c}{Agent 1 dose level} \\
Agent 2 & 1 & 2 & 3 & 4 & 5 \\ \hline
3 & 0.20 & 0.30 & 0.40 & 0.55 & 0.68 \\
2 & 0.15 & 0.25 & 0.35 & 0.50 & 0.63 \\
1 & 0.10 & 0.20 & 0.30 & 0.45 & 0.58 \\
\hline\end{tabular}\end{minipage}
 \\[10pt]
\multicolumn{2}{c}{\textbf{Scenario 4}}\\*[3pt]
\begin{minipage}[t]{\linewidth}\centering
\textbf{Toxicity probability}\par\smallskip
\begin{tabular}{@{}rrrrrr@{}}
 & \multicolumn{5}{c}{Agent 1 dose level} \\
Agent 2 & 1 & 2 & 3 & 4 & 5 \\ \hline
3 & 0.50 & 0.60 & 0.70 & 0.80 & 0.90 \\
2 & 0.45 & 0.55 & 0.65 & 0.75 & 0.85 \\
1 & \textbf{0.30} & 0.45 & 0.60 & 0.70 & 0.80 \\
\hline\end{tabular}\end{minipage}
 & \begin{minipage}[t]{\linewidth}\centering
\textbf{Efficacy probability}\par\smallskip
\begin{tabular}{@{}rrrrrr@{}}
 & \multicolumn{5}{c}{Agent 1 dose level} \\
Agent 2 & 1 & 2 & 3 & 4 & 5 \\ \hline
3 & 0.15 & 0.20 & 0.25 & 0.40 & 0.55 \\
2 & 0.10 & 0.15 & 0.20 & 0.35 & 0.50 \\
1 & 0.05 & 0.10 & 0.15 & 0.30 & 0.45 \\
\hline\end{tabular}\end{minipage}
 \\[10pt]
\multicolumn{2}{c}{\textbf{Scenario 5}}\\*[3pt]
\begin{minipage}[t]{\linewidth}\centering
\textbf{Toxicity probability}\par\smallskip
\begin{tabular}{@{}rrrrrr@{}}
 & \multicolumn{5}{c}{Agent 1 dose level} \\
Agent 2 & 1 & 2 & 3 & 4 & 5 \\ \hline
3 & 0.07 & 0.09 & 0.12 & 0.15 & \textbf{0.30} \\
2 & 0.03 & 0.05 & 0.10 & 0.13 & 0.15 \\
1 & 0.01 & 0.02 & 0.08 & 0.10 & 0.11 \\
\hline\end{tabular}\end{minipage}
 & \begin{minipage}[t]{\linewidth}\centering
\textbf{Efficacy probability}\par\smallskip
\begin{tabular}{@{}rrrrrr@{}}
 & \multicolumn{5}{c}{Agent 1 dose level} \\
Agent 2 & 1 & 2 & 3 & 4 & 5 \\ \hline
3 & 0.15 & 0.20 & 0.25 & 0.30 & 0.40 \\
2 & 0.10 & 0.15 & 0.20 & 0.25 & 0.35 \\
1 & 0.05 & 0.10 & 0.15 & 0.20 & 0.30 \\
\hline\end{tabular}\end{minipage}
 \\[10pt]
\multicolumn{2}{c}{\textbf{Scenario 6}}\\*[3pt]
\begin{minipage}[t]{\linewidth}\centering
\textbf{Toxicity probability}\par\smallskip
\begin{tabular}{@{}rrrrrr@{}}
 & \multicolumn{5}{c}{Agent 1 dose level} \\
Agent 2 & 1 & 2 & 3 & 4 & 5 \\ \hline
3 & 0.15 & \textbf{0.30} & 0.45 & 0.50 & 0.60 \\
2 & 0.09 & 0.12 & 0.15 & \textbf{0.30} & 0.45 \\
1 & 0.05 & 0.08 & 0.10 & 0.13 & 0.15 \\
\hline\end{tabular}\end{minipage}
 & \begin{minipage}[t]{\linewidth}\centering
\textbf{Efficacy probability}\par\smallskip
\begin{tabular}{@{}rrrrrr@{}}
 & \multicolumn{5}{c}{Agent 1 dose level} \\
Agent 2 & 1 & 2 & 3 & 4 & 5 \\ \hline
3 & 0.40 & 0.45 & 0.46 & 0.46 & 0.46 \\
2 & 0.35 & 0.40 & 0.41 & 0.41 & 0.41 \\
1 & 0.30 & 0.35 & 0.36 & 0.36 & 0.36 \\
\hline\end{tabular}\end{minipage}
 \\[10pt]
\multicolumn{2}{c}{\textbf{Scenario 7}}\\*[3pt]
\begin{minipage}[t]{\linewidth}\centering
\textbf{Toxicity probability}\par\smallskip
\begin{tabular}{@{}rrrrrr@{}}
 & \multicolumn{5}{c}{Agent 1 dose level} \\
Agent 2 & 1 & 2 & 3 & 4 & 5 \\ \hline
3 & \textbf{0.30} & 0.50 & 0.60 & 0.65 & 0.75 \\
2 & 0.15 & \textbf{0.30} & 0.45 & 0.52 & 0.60 \\
1 & 0.07 & 0.10 & 0.12 & 0.15 & \textbf{0.30} \\
\hline\end{tabular}\end{minipage}
 & \begin{minipage}[t]{\linewidth}\centering
\textbf{Efficacy probability}\par\smallskip
\begin{tabular}{@{}rrrrrr@{}}
 & \multicolumn{5}{c}{Agent 1 dose level} \\
Agent 2 & 1 & 2 & 3 & 4 & 5 \\ \hline
3 & 0.25 & 0.40 & 0.45 & 0.46 & 0.46 \\
2 & 0.20 & 0.35 & 0.40 & 0.41 & 0.41 \\
1 & 0.15 & 0.30 & 0.35 & 0.36 & 0.36 \\
\hline\end{tabular}\end{minipage}
 \\[10pt]
\multicolumn{2}{c}{\textbf{Scenario 8}}\\*[3pt]
\begin{minipage}[t]{\linewidth}\centering
\textbf{Toxicity probability}\par\smallskip
\begin{tabular}{@{}rrrrrr@{}}
 & \multicolumn{5}{c}{Agent 1 dose level} \\
Agent 2 & 1 & 2 & 3 & 4 & 5 \\ \hline
3 & 0.08 & 0.15 & 0.45 & 0.60 & 0.80 \\
2 & 0.05 & 0.12 & \textbf{0.30} & 0.55 & 0.70 \\
1 & 0.02 & 0.10 & 0.15 & 0.50 & 0.60 \\
\hline\end{tabular}\end{minipage}
 & \begin{minipage}[t]{\linewidth}\centering
\textbf{Efficacy probability}\par\smallskip
\begin{tabular}{@{}rrrrrr@{}}
 & \multicolumn{5}{c}{Agent 1 dose level} \\
Agent 2 & 1 & 2 & 3 & 4 & 5 \\ \hline
3 & 0.20 & 0.30 & 0.40 & 0.45 & 0.45 \\
2 & 0.15 & 0.25 & 0.35 & 0.40 & 0.40 \\
1 & 0.10 & 0.20 & 0.30 & 0.35 & 0.35 \\
\hline\end{tabular}\end{minipage}
 \\[10pt]
\multicolumn{2}{c}{\textbf{Scenario 9}}\\*[3pt]
\begin{minipage}[t]{\linewidth}\centering
\textbf{Toxicity probability}\par\smallskip
\begin{tabular}{@{}rrrrrr@{}}
 & \multicolumn{5}{c}{Agent 1 dose level} \\
Agent 2 & 1 & 2 & 3 & 4 & 5 \\ \hline
3 & 0.15 & \textbf{0.30} & 0.45 & 0.55 & 0.65 \\
2 & 0.02 & 0.05 & 0.08 & 0.12 & 0.15 \\
1 & 0.005 & 0.01 & 0.02 & 0.04 & 0.07 \\
\hline\end{tabular}\end{minipage}
 & \begin{minipage}[t]{\linewidth}\centering
\textbf{Efficacy probability}\par\smallskip
\begin{tabular}{@{}rrrrrr@{}}
 & \multicolumn{5}{c}{Agent 1 dose level} \\
Agent 2 & 1 & 2 & 3 & 4 & 5 \\ \hline
3 & 0.20 & 0.25 & 0.30 & 0.40 & 0.45 \\
2 & 0.15 & 0.20 & 0.25 & 0.35 & 0.40 \\
1 & 0.10 & 0.15 & 0.20 & 0.30 & 0.35 \\
\hline\end{tabular}\end{minipage}
 \\[10pt]
\multicolumn{2}{c}{\textbf{Scenario 10}}\\*[3pt]
\begin{minipage}[t]{\linewidth}\centering
\textbf{Toxicity probability}\par\smallskip
\begin{tabular}{@{}rrrrrr@{}}
 & \multicolumn{5}{c}{Agent 1 dose level} \\
Agent 2 & 1 & 2 & 3 & 4 & 5 \\ \hline
3 & 0.70 & 0.75 & 0.80 & 0.85 & 0.90 \\
2 & 0.45 & 0.50 & 0.60 & 0.65 & 0.70 \\
1 & 0.05 & 0.10 & 0.15 & \textbf{0.30} & 0.45 \\
\hline\end{tabular}\end{minipage}
 & \begin{minipage}[t]{\linewidth}\centering
\textbf{Efficacy probability}\par\smallskip
\begin{tabular}{@{}rrrrrr@{}}
 & \multicolumn{5}{c}{Agent 1 dose level} \\
Agent 2 & 1 & 2 & 3 & 4 & 5 \\ \hline
3 & 0.40 & 0.42 & 0.45 & 0.46 & 0.46 \\
2 & 0.35 & 0.37 & 0.40 & 0.41 & 0.41 \\
1 & 0.30 & 0.32 & 0.35 & 0.36 & 0.36 \\
\hline\end{tabular}\end{minipage}
 \\[10pt]
\multicolumn{2}{c}{\textbf{Scenario 11}}\\*[3pt]
\begin{minipage}[t]{\linewidth}\centering
\textbf{Toxicity probability}\par\smallskip
\begin{tabular}{@{}rrrr@{}}
 & \multicolumn{3}{c}{Agent 1 dose level} \\
Agent 2 & 1 & 2 & 3 \\ \hline
3 & 0.08 & 0.10 & \textbf{0.18} \\
2 & 0.04 & 0.06 & 0.08 \\
1 & 0.02 & 0.04 & 0.06 \\
\hline\end{tabular}\end{minipage}
 & \begin{minipage}[t]{\linewidth}\centering
\textbf{Efficacy probability}\par\smallskip
\begin{tabular}{@{}rrrr@{}}
 & \multicolumn{3}{c}{Agent 1 dose level} \\
Agent 2 & 1 & 2 & 3 \\ \hline
3 & 0.15 & 0.20 & 0.40 \\
2 & 0.10 & 0.16 & 0.20 \\
1 & 0.05 & 0.10 & 0.15 \\
\hline\end{tabular}\end{minipage}
 \\[10pt]
\multicolumn{2}{c}{\textbf{Scenario 12}}\\*[3pt]
\begin{minipage}[t]{\linewidth}\centering
\textbf{Toxicity probability}\par\smallskip
\begin{tabular}{@{}rrrr@{}}
 & \multicolumn{3}{c}{Agent 1 dose level} \\
Agent 2 & 1 & 2 & 3 \\ \hline
3 & 0.16 & \textbf{0.25} & \textbf{0.35} \\
2 & 0.10 & 0.14 & 0.20 \\
1 & 0.06 & 0.08 & 0.12 \\
\hline\end{tabular}\end{minipage}
 & \begin{minipage}[t]{\linewidth}\centering
\textbf{Efficacy probability}\par\smallskip
\begin{tabular}{@{}rrrr@{}}
 & \multicolumn{3}{c}{Agent 1 dose level} \\
Agent 2 & 1 & 2 & 3 \\ \hline
3 & 0.20 & 0.35 & 0.50 \\
2 & 0.10 & 0.25 & 0.40 \\
1 & 0.05 & 0.10 & 0.20 \\
\hline\end{tabular}\end{minipage}
 \\[10pt]
\multicolumn{2}{c}{\textbf{Scenario 13}}\\*[3pt]
\begin{minipage}[t]{\linewidth}\centering
\textbf{Toxicity probability}\par\smallskip
\begin{tabular}{@{}rrrr@{}}
 & \multicolumn{3}{c}{Agent 1 dose level} \\
Agent 2 & 1 & 2 & 3 \\ \hline
3 & 0.24 & \textbf{0.33} & 0.40 \\
2 & 0.16 & 0.22 & 0.35 \\
1 & 0.08 & 0.14 & 0.20 \\
\hline\end{tabular}\end{minipage}
 & \begin{minipage}[t]{\linewidth}\centering
\textbf{Efficacy probability}\par\smallskip
\begin{tabular}{@{}rrrr@{}}
 & \multicolumn{3}{c}{Agent 1 dose level} \\
Agent 2 & 1 & 2 & 3 \\ \hline
3 & 0.40 & 0.50 & 0.60 \\
2 & 0.20 & 0.40 & 0.50 \\
1 & 0.10 & 0.25 & 0.35 \\
\hline\end{tabular}\end{minipage}
 \\[10pt]
\multicolumn{2}{c}{\textbf{Scenario 14}}\\*[3pt]
\begin{minipage}[t]{\linewidth}\centering
\textbf{Toxicity probability}\par\smallskip
\begin{tabular}{@{}rrrr@{}}
 & \multicolumn{3}{c}{Agent 1 dose level} \\
Agent 2 & 1 & 2 & 3 \\ \hline
3 & \textbf{0.33} & 0.40 & 0.55 \\
2 & 0.18 & 0.25 & 0.42 \\
1 & 0.12 & 0.20 & 0.35 \\
\hline\end{tabular}\end{minipage}
 & \begin{minipage}[t]{\linewidth}\centering
\textbf{Efficacy probability}\par\smallskip
\begin{tabular}{@{}rrrr@{}}
 & \multicolumn{3}{c}{Agent 1 dose level} \\
Agent 2 & 1 & 2 & 3 \\ \hline
3 & 0.50 & 0.60 & 0.70 \\
2 & 0.35 & 0.45 & 0.55 \\
1 & 0.20 & 0.40 & 0.50 \\
\hline\end{tabular}\end{minipage}
 \\[10pt]
\multicolumn{2}{c}{\textbf{Scenario 15}}\\*[3pt]
\begin{minipage}[t]{\linewidth}\centering
\textbf{Toxicity probability}\par\smallskip
\begin{tabular}{@{}rrrr@{}}
 & \multicolumn{3}{c}{Agent 1 dose level} \\
Agent 2 & 1 & 2 & 3 \\ \hline
3 & 0.45 & 0.55 & 0.75 \\
2 & 0.20 & \textbf{0.35} & 0.40 \\
1 & 0.15 & 0.20 & \textbf{0.25} \\
\hline\end{tabular}\end{minipage}
 & \begin{minipage}[t]{\linewidth}\centering
\textbf{Efficacy probability}\par\smallskip
\begin{tabular}{@{}rrrr@{}}
 & \multicolumn{3}{c}{Agent 1 dose level} \\
Agent 2 & 1 & 2 & 3 \\ \hline
3 & 0.55 & 0.65 & 0.75 \\
2 & 0.36 & 0.49 & 0.62 \\
1 & 0.20 & 0.35 & 0.50 \\
\hline\end{tabular}\end{minipage}
 \\[10pt]
\multicolumn{2}{c}{\textbf{Scenario 16}}\\*[3pt]
\begin{minipage}[t]{\linewidth}\centering
\textbf{Toxicity probability}\par\smallskip
\begin{tabular}{@{}rrrr@{}}
 & \multicolumn{3}{c}{Agent 1 dose level} \\
Agent 2 & 1 & 2 & 3 \\ \hline
3 & 0.65 & 0.80 & 0.85 \\
2 & 0.55 & 0.70 & 0.75 \\
1 & 0.50 & 0.55 & 0.65 \\
\hline\end{tabular}\end{minipage}
 & \begin{minipage}[t]{\linewidth}\centering
\textbf{Efficacy probability}\par\smallskip
\begin{tabular}{@{}rrrr@{}}
 & \multicolumn{3}{c}{Agent 1 dose level} \\
Agent 2 & 1 & 2 & 3 \\ \hline
3 & 0.60 & 0.65 & 0.70 \\
2 & 0.55 & 0.60 & 0.65 \\
1 & 0.50 & 0.55 & 0.60 \\
\hline\end{tabular}\end{minipage}
 \\[10pt]
\end{longtable}\endgroup

%% file: main_sim_results.tex
% Add \usepackage{booktabs,graphicx,pdflscape} to your main preamble.
% Input this file after the references, before \end{document}.
% Requires citation key cheung2011dose in your bibliography.
\subsection{Main Simulation Results}
The tables in the subsequent pages provide results from the simulation studies.
\label{app:sim_result}
\begin{landscape}
\begin{table}[p]
\centering
\caption{Operating characteristics of BF-BOIN-Comb and BOIN-Comb with 10 cohorts (maximum escalation enrollment 30 patients), $n_{\mathrm{stop}}=9$, and 5,000 simulated trials per scenario and design. AI denotes the accuracy index\cite{cheung2011dose}.}
\label{tab:oc-30}
\scriptsize\setlength{\tabcolsep}{4pt}\renewcommand{\arraystretch}{1.0}
\resizebox{0.92\linewidth}{!}{%
\begin{tabular}{@{}rlrrrrrrrr@{}}
\toprule
\shortstack{Scenario} & Method & \shortstack{Correct\\selection\\(\%)} & \shortstack{AI} & \shortstack{Overdosing\\selection\\(\%)} & \shortstack{Patients\\below MTD\\$n$ (\%)} & \shortstack{Patients\\at MTD(s)\\$n$ (\%)} & \shortstack{Patients\\above MTD\\$n$ (\%)} & \shortstack{Average\\sample size\\($n$)} & \shortstack{Average\\duration\\(months)}\\
\midrule
1 & BF-BOIN-Comb & 52.9 & 0.516 & 11.4 & 29.8 (70.8\%) & 9.4 (21.2\%) & 3.8 (8.0\%) & 43.0 & 16.6 \\
 & BOIN-Comb & 52.5 & 0.500 & 24.5 & 14.3 (54.8\%) & 7.7 (28.0\%) & 5.0 (17.2\%) & 27.0 & 16.1 \\
\addlinespace[1pt]
2 & BF-BOIN-Comb & 61.3 & 0.752 & 20.3 & 10.1 (41.5\%) & 11.7 (39.5\%) & 6.5 (18.9\%) & 28.4 & 13.2 \\
 & BOIN-Comb & 52.9 & 0.686 & 29.4 & 6.6 (34.6\%) & 9.0 (38.3\%) & 7.2 (27.0\%) & 22.8 & 13.4 \\
\addlinespace[1pt]
3 & BF-BOIN-Comb & 50.1 & 0.487 & 11.1 & 31.6 (74.9\%) & 8.1 (18.5\%) & 3.0 (6.6\%) & 42.7 & 17.4 \\
 & BOIN-Comb & 50.4 & 0.485 & 21.9 & 17.4 (63.3\%) & 6.8 (23.8\%) & 3.8 (12.9\%) & 27.9 & 16.8 \\
\addlinespace[1pt]
4 & BF-BOIN-Comb & 59.2 & 0.850 & 28.2 & 0.0 (0.0\%) & 9.5 (69.1\%) & 7.6 (30.9\%) & 17.1 & 9.3 \\
 & BOIN-Comb & 58.5 & 0.842 & 29.0 & 0.0 (0.0\%) & 8.9 (67.4\%) & 7.6 (32.6\%) & 16.5 & 9.5 \\
\addlinespace[1pt]
5 & BF-BOIN-Comb & 45.2 & 0.551 & 0.0 & 38.9 (89.5\%) & 4.6 (10.5\%) & 0.0 (0.0\%) & 43.6 & 18.2 \\
 & BOIN-Comb & 57.8 & 0.658 & 0.0 & 23.7 (82.0\%) & 5.3 (18.0\%) & 0.0 (0.0\%) & 28.9 & 17.6 \\
\addlinespace[1pt]
6 & BF-BOIN-Comb & 40.5 & 0.413 & 14.5 & 33.7 (76.4\%) & 6.6 (14.6\%) & 4.3 (9.0\%) & 44.6 & 17.1 \\
 & BOIN-Comb & 42.3 & 0.430 & 26.8 & 16.5 (61.1\%) & 5.7 (20.8\%) & 5.3 (18.1\%) & 27.5 & 16.5 \\
\addlinespace[1pt]
7 & BF-BOIN-Comb & 54.7 & 0.602 & 15.5 & 22.8 (61.4\%) & 9.0 (24.8\%) & 5.6 (13.8\%) & 37.4 & 15.8 \\
 & BOIN-Comb & 51.8 & 0.560 & 23.0 & 12.8 (50.4\%) & 7.1 (27.7\%) & 6.1 (21.9\%) & 26.0 & 15.5 \\
\addlinespace[1pt]
8 & BF-BOIN-Comb & 30.0 & 0.510 & 23.4 & 28.9 (72.5\%) & 4.8 (11.7\%) & 6.6 (15.9\%) & 40.4 & 16.8 \\
 & BOIN-Comb & 25.9 & 0.476 & 33.2 & 16.6 (62.2\%) & 3.4 (12.4\%) & 7.2 (25.4\%) & 27.3 & 16.3 \\
\addlinespace[1pt]
9 & BF-BOIN-Comb & 32.0 & 0.479 & 19.4 & 31.2 (72.8\%) & 5.1 (13.4\%) & 5.9 (13.8\%) & 42.3 & 17.3 \\
 & BOIN-Comb & 26.3 & 0.425 & 25.7 & 18.0 (64.4\%) & 3.7 (14.4\%) & 6.0 (21.1\%) & 27.8 & 16.7 \\
\addlinespace[1pt]
10 & BF-BOIN-Comb & 22.2 & 0.584 & 42.4 & 24.1 (62.5\%) & 3.0 (6.7\%) & 11.1 (30.8\%) & 38.1 & 16.1 \\
 & BOIN-Comb & 25.6 & 0.587 & 43.9 & 13.5 (50.1\%) & 2.9 (10.1\%) & 10.3 (39.8\%) & 26.8 & 15.8 \\
\addlinespace[1pt]
11 & BF-BOIN-Comb & 89.5 & 0.427 & 0.0 & 25.5 (73.9\%) & 8.5 (26.1\%) & 0.0 (0.0\%) & 34.0 & 15.1 \\
 & BOIN-Comb & 91.4 & 0.435 & 0.0 & 14.9 (63.0\%) & 8.6 (37.0\%) & 0.0 (0.0\%) & 23.5 & 14.3 \\
\addlinespace[1pt]
12 & BF-BOIN-Comb & 54.7 & 0.413 & 0.0 & 26.1 (73.4\%) & 9.8 (26.6\%) & 0.0 (0.0\%) & 35.9 & 16.0 \\
 & BOIN-Comb & 62.5 & 0.465 & 0.0 & 16.7 (66.0\%) & 9.0 (34.0\%) & 0.0 (0.0\%) & 25.7 & 15.5 \\
\addlinespace[1pt]
13 & BF-BOIN-Comb & 16.1 & 0.228 & 22.6 & 27.2 (78.1\%) & 3.3 (8.5\%) & 5.3 (13.4\%) & 35.9 & 15.5 \\
 & BOIN-Comb & 17.4 & 0.233 & 34.8 & 16.5 (68.0\%) & 2.9 (10.7\%) & 5.7 (21.3\%) & 25.1 & 15.0 \\
\addlinespace[1pt]
14 & BF-BOIN-Comb & 14.5 & 0.260 & 27.4 & 25.1 (76.4\%) & 2.4 (6.6\%) & 6.7 (16.9\%) & 34.2 & 14.6 \\
 & BOIN-Comb & 12.6 & 0.197 & 38.1 & 15.1 (65.4\%) & 1.9 (7.5\%) & 7.3 (27.1\%) & 24.4 & 14.4 \\
\addlinespace[1pt]
15 & BF-BOIN-Comb & 34.3 & 0.387 & 16.5 & 20.2 (69.3\%) & 6.6 (18.0\%) & 4.9 (12.7\%) & 31.7 & 13.9 \\
 & BOIN-Comb & 33.9 & 0.372 & 25.6 & 12.3 (58.3\%) & 5.4 (20.6\%) & 5.7 (21.1\%) & 23.4 & 13.8 \\
\bottomrule
\end{tabular}%
}
\par\smallskip
\begin{minipage}{\linewidth}
\scriptsize Counts and percentages are averaged across trials. AI is calculated among trials where a dose is selected (i.e. trials ending due to excessive toxicity are excluded); Percent correct selection  includes all trials, with no selection counted as a failure. Below and above MTD refer to dose combinations with true toxicity probabilities strictly below the smallest and strictly above the largest true MTD toxicity probability, respectively. Average sample size includes escalation and backfill patients; the escalation limit does not cap total BF-BOIN-Comb enrollment.
\end{minipage}
\end{table}
\end{landscape}
\clearpage
\begin{landscape}
\begin{table}[p]
\centering
\caption{Operating characteristics of BF-BOIN-Comb and BOIN-Comb with 15 cohorts (maximum escalation enrollment 45 patients), $n_{\mathrm{stop}}=9$, and 5,000 simulated trials per scenario and design. AI denotes the accuracy index\cite{cheung2011dose}.}
\label{tab:oc-45}
\scriptsize\setlength{\tabcolsep}{4pt}\renewcommand{\arraystretch}{1.0}
\resizebox{0.92\linewidth}{!}{%
\begin{tabular}{@{}rlrrrrrrrr@{}}
\toprule
\shortstack{Scenario} & Method & \shortstack{Correct\\selection\\(\%)} & \shortstack{AI} & \shortstack{Overdosing\\selection\\(\%)} & \shortstack{Patients\\below MTD\\$n$ (\%)} & \shortstack{Patients\\at MTD(s)\\$n$ (\%)} & \shortstack{Patients\\above MTD\\$n$ (\%)} & \shortstack{Average\\sample size\\($n$)} & \shortstack{Average\\duration\\(months)}\\
\midrule
1 & BF-BOIN-Comb & 54.8 & 0.535 & 11.0 & 33.8 (69.5\%) & 11.6 (21.8\%) & 5.1 (8.7\%) & 50.5 & 19.4 \\
 & BOIN-Comb & 54.6 & 0.527 & 24.5 & 15.0 (50.8\%) & 9.5 (29.4\%) & 7.0 (19.7\%) & 31.4 & 18.7 \\
\addlinespace[1pt]
2 & BF-BOIN-Comb & 61.9 & 0.756 & 20.2 & 10.8 (40.8\%) & 13.1 (39.9\%) & 7.5 (19.4\%) & 31.4 & 14.8 \\
 & BOIN-Comb & 55.3 & 0.703 & 28.5 & 7.0 (33.6\%) & 10.2 (38.4\%) & 8.6 (28.0\%) & 25.8 & 15.1 \\
\addlinespace[1pt]
3 & BF-BOIN-Comb & 55.1 & 0.540 & 10.2 & 36.5 (71.6\%) & 11.0 (20.4\%) & 4.6 (8.0\%) & 52.2 & 20.9 \\
 & BOIN-Comb & 53.8 & 0.522 & 22.2 & 18.2 (57.8\%) & 8.9 (25.9\%) & 6.0 (16.3\%) & 33.1 & 19.8 \\
\addlinespace[1pt]
4 & BF-BOIN-Comb & 61.1 & 0.861 & 26.3 & 0.0 (0.0\%) & 9.9 (69.6\%) & 8.2 (30.4\%) & 18.1 & 9.9 \\
 & BOIN-Comb & 58.1 & 0.841 & 28.9 & 0.0 (0.0\%) & 9.2 (67.1\%) & 8.4 (32.9\%) & 17.6 & 10.1 \\
\addlinespace[1pt]
5 & BF-BOIN-Comb & 53.7 & 0.620 & 0.0 & 44.7 (86.9\%) & 6.8 (13.1\%) & 0.0 (0.0\%) & 51.4 & 21.0 \\
 & BOIN-Comb & 66.6 & 0.728 & 0.0 & 25.3 (77.4\%) & 7.5 (22.6\%) & 0.0 (0.0\%) & 32.9 & 20.0 \\
\addlinespace[1pt]
6 & BF-BOIN-Comb & 43.1 & 0.446 & 14.3 & 39.3 (74.1\%) & 8.7 (15.8\%) & 6.0 (10.1\%) & 54.0 & 20.4 \\
 & BOIN-Comb & 44.0 & 0.450 & 27.5 & 17.9 (56.5\%) & 7.2 (21.8\%) & 7.8 (21.7\%) & 32.9 & 19.7 \\
\addlinespace[1pt]
7 & BF-BOIN-Comb & 60.2 & 0.650 & 13.8 & 25.9 (59.6\%) & 11.3 (26.2\%) & 7.0 (14.2\%) & 44.1 & 18.7 \\
 & BOIN-Comb & 57.2 & 0.611 & 21.7 & 14.0 (47.3\%) & 8.9 (29.5\%) & 7.8 (23.2\%) & 30.7 & 18.2 \\
\addlinespace[1pt]
8 & BF-BOIN-Comb & 33.1 & 0.536 & 20.6 & 33.3 (71.0\%) & 6.3 (12.8\%) & 8.3 (16.3\%) & 48.0 & 19.9 \\
 & BOIN-Comb & 27.9 & 0.496 & 32.1 & 18.2 (59.4\%) & 4.7 (13.8\%) & 9.2 (26.9\%) & 32.1 & 19.1 \\
\addlinespace[1pt]
9 & BF-BOIN-Comb & 35.0 & 0.525 & 17.8 & 37.5 (71.7\%) & 6.3 (14.0\%) & 7.6 (14.2\%) & 51.5 & 20.7 \\
 & BOIN-Comb & 28.2 & 0.467 & 23.7 & 20.6 (62.3\%) & 4.4 (15.0\%) & 7.9 (22.7\%) & 33.0 & 19.8 \\
\addlinespace[1pt]
10 & BF-BOIN-Comb & 31.1 & 0.636 & 40.6 & 28.2 (60.1\%) & 5.2 (9.5\%) & 13.6 (30.4\%) & 47.0 & 20.1 \\
 & BOIN-Comb & 31.2 & 0.626 & 44.1 & 15.4 (46.9\%) & 4.6 (12.6\%) & 12.8 (40.5\%) & 32.8 & 19.4 \\
\addlinespace[1pt]
11 & BF-BOIN-Comb & 91.1 & 0.434 & 0.0 & 26.1 (73.7\%) & 8.7 (26.3\%) & 0.0 (0.0\%) & 34.8 & 15.4 \\
 & BOIN-Comb & 92.6 & 0.439 & 0.0 & 15.1 (62.8\%) & 8.8 (37.2\%) & 0.0 (0.0\%) & 23.8 & 14.5 \\
\addlinespace[1pt]
12 & BF-BOIN-Comb & 56.3 & 0.419 & 0.0 & 28.5 (72.8\%) & 11.2 (27.2\%) & 0.0 (0.0\%) & 39.7 & 17.6 \\
 & BOIN-Comb & 64.7 & 0.470 & 0.0 & 17.5 (64.6\%) & 10.2 (35.4\%) & 0.0 (0.0\%) & 27.7 & 16.7 \\
\addlinespace[1pt]
13 & BF-BOIN-Comb & 15.3 & 0.222 & 22.4 & 30.2 (77.6\%) & 3.9 (8.7\%) & 6.3 (13.7\%) & 40.3 & 17.4 \\
 & BOIN-Comb & 18.7 & 0.244 & 34.3 & 17.6 (66.2\%) & 3.5 (11.4\%) & 6.8 (22.3\%) & 27.9 & 16.7 \\
\addlinespace[1pt]
14 & BF-BOIN-Comb & 14.6 & 0.268 & 28.4 & 27.5 (75.6\%) & 2.9 (6.9\%) & 8.0 (17.6\%) & 38.4 & 16.5 \\
 & BOIN-Comb & 13.9 & 0.217 & 40.1 & 16.1 (63.1\%) & 2.3 (8.0\%) & 9.0 (28.9\%) & 27.4 & 16.3 \\
\addlinespace[1pt]
15 & BF-BOIN-Comb & 35.6 & 0.396 & 16.8 & 22.1 (68.3\%) & 7.8 (18.9\%) & 5.6 (12.8\%) & 35.5 & 15.7 \\
 & BOIN-Comb & 35.7 & 0.385 & 25.1 & 13.2 (56.6\%) & 6.5 (21.9\%) & 6.7 (21.6\%) & 26.3 & 15.6 \\
\bottomrule
\end{tabular}%
}
\par\smallskip
\begin{minipage}{\linewidth}
\scriptsize Counts and percentages are averaged across trials. AI is calculated among trials where a dose is selected (i.e. trials ending due to excessive toxicity are excluded); Percent correct selection includes all trials, with no selection counted as a failure. Below and above MTD refer to dose combinations with true toxicity probabilities strictly below the smallest and strictly above the largest true MTD toxicity probability, respectively. Average sample size includes escalation and backfill patients; the escalation limit does not cap total BF-BOIN-Comb enrollment.
\end{minipage}
\end{table}
\end{landscape}
\clearpage
\begin{landscape}
\begin{table}[p]
\centering
\caption{Operating characteristics of BF-BOIN-Comb and BOIN-Comb with 20 cohorts (maximum escalation enrollment 60 patients), $n_{\mathrm{stop}}=9$, and 5,000 simulated trials per scenario and design. AI denotes the accuracy index\cite{cheung2011dose}.}
\label{tab:oc-60}
\scriptsize\setlength{\tabcolsep}{4pt}\renewcommand{\arraystretch}{1.0}
\resizebox{0.92\linewidth}{!}{%
\begin{tabular}{@{}rlrrrrrrrr@{}}
\toprule
\shortstack{Scenario} & Method & \shortstack{Correct\\selection\\(\%)} & \shortstack{AI} & \shortstack{Overdosing\\selection\\(\%)} & \shortstack{Patients\\below MTD\\$n$ (\%)} & \shortstack{Patients\\at MTD(s)\\$n$ (\%)} & \shortstack{Patients\\above MTD\\$n$ (\%)} & \shortstack{Average\\sample size\\($n$)} & \shortstack{Average\\duration\\(months)}\\
\midrule
1 & BF-BOIN-Comb & 56.2 & 0.550 & 10.7 & 34.9 (68.8\%) & 12.6 (22.3\%) & 5.5 (8.9\%) & 52.9 & 20.4 \\
 & BOIN-Comb & 55.1 & 0.531 & 24.6 & 15.3 (50.0\%) & 10.2 (29.9\%) & 7.5 (20.1\%) & 33.0 & 19.7 \\
\addlinespace[1pt]
2 & BF-BOIN-Comb & 62.9 & 0.762 & 19.8 & 11.1 (40.7\%) & 13.7 (40.0\%) & 7.8 (19.3\%) & 32.6 & 15.5 \\
 & BOIN-Comb & 55.4 & 0.704 & 28.6 & 7.1 (33.3\%) & 10.6 (38.7\%) & 8.8 (28.0\%) & 26.5 & 15.6 \\
\addlinespace[1pt]
3 & BF-BOIN-Comb & 55.5 & 0.545 & 10.2 & 38.0 (71.3\%) & 12.0 (20.7\%) & 4.9 (7.9\%) & 54.9 & 21.9 \\
 & BOIN-Comb & 54.4 & 0.529 & 21.3 & 18.6 (57.5\%) & 9.6 (26.5\%) & 6.2 (16.1\%) & 34.3 & 20.6 \\
\addlinespace[1pt]
4 & BF-BOIN-Comb & 61.0 & 0.860 & 26.4 & 0.0 (0.0\%) & 10.0 (69.3\%) & 8.4 (30.7\%) & 18.5 & 10.0 \\
 & BOIN-Comb & 58.3 & 0.843 & 28.6 & 0.0 (0.0\%) & 9.4 (67.4\%) & 8.5 (32.6\%) & 17.8 & 10.3 \\
\addlinespace[1pt]
5 & BF-BOIN-Comb & 55.2 & 0.634 & 0.0 & 45.8 (86.5\%) & 7.2 (13.5\%) & 0.0 (0.0\%) & 52.9 & 21.6 \\
 & BOIN-Comb & 65.7 & 0.721 & 0.0 & 25.7 (77.4\%) & 7.7 (22.6\%) & 0.0 (0.0\%) & 33.5 & 20.4 \\
\addlinespace[1pt]
6 & BF-BOIN-Comb & 43.1 & 0.443 & 15.2 & 41.2 (73.9\%) & 9.3 (15.9\%) & 6.6 (10.3\%) & 57.1 & 21.6 \\
 & BOIN-Comb & 45.5 & 0.466 & 26.1 & 18.4 (56.4\%) & 7.8 (22.4\%) & 8.0 (21.2\%) & 34.2 & 20.5 \\
\addlinespace[1pt]
7 & BF-BOIN-Comb & 63.4 & 0.679 & 13.0 & 26.7 (59.2\%) & 12.1 (26.8\%) & 7.1 (14.0\%) & 45.9 & 19.6 \\
 & BOIN-Comb & 59.3 & 0.631 & 21.5 & 14.3 (46.7\%) & 9.5 (30.0\%) & 8.1 (23.3\%) & 32.0 & 19.0 \\
\addlinespace[1pt]
8 & BF-BOIN-Comb & 33.8 & 0.543 & 21.2 & 34.8 (70.5\%) & 7.0 (13.0\%) & 9.0 (16.4\%) & 50.8 & 21.2 \\
 & BOIN-Comb & 29.2 & 0.504 & 31.6 & 18.6 (59.2\%) & 5.0 (13.7\%) & 9.8 (27.1\%) & 33.3 & 19.9 \\
\addlinespace[1pt]
9 & BF-BOIN-Comb & 39.4 & 0.557 & 16.6 & 38.5 (70.8\%) & 7.2 (15.1\%) & 7.9 (14.1\%) & 53.6 & 21.7 \\
 & BOIN-Comb & 29.2 & 0.477 & 22.1 & 21.4 (62.5\%) & 4.6 (15.0\%) & 8.2 (22.4\%) & 34.2 & 20.5 \\
\addlinespace[1pt]
10 & BF-BOIN-Comb & 33.5 & 0.653 & 39.3 & 29.8 (59.6\%) & 6.1 (10.3\%) & 14.4 (30.1\%) & 50.3 & 21.8 \\
 & BOIN-Comb & 34.9 & 0.651 & 43.0 & 16.3 (46.2\%) & 5.6 (13.9\%) & 13.6 (39.9\%) & 35.5 & 21.0 \\
\addlinespace[1pt]
11 & BF-BOIN-Comb & 91.4 & 0.435 & 0.0 & 26.1 (73.6\%) & 8.8 (26.4\%) & 0.0 (0.0\%) & 34.9 & 15.4 \\
 & BOIN-Comb & 92.8 & 0.440 & 0.0 & 15.1 (62.8\%) & 8.8 (37.2\%) & 0.0 (0.0\%) & 23.9 & 14.6 \\
\addlinespace[1pt]
12 & BF-BOIN-Comb & 57.7 & 0.432 & 0.0 & 29.3 (72.4\%) & 11.6 (27.6\%) & 0.0 (0.0\%) & 40.9 & 18.1 \\
 & BOIN-Comb & 64.9 & 0.472 & 0.0 & 17.7 (64.4\%) & 10.4 (35.6\%) & 0.0 (0.0\%) & 28.1 & 16.9 \\
\addlinespace[1pt]
13 & BF-BOIN-Comb & 15.4 & 0.228 & 23.6 & 30.7 (77.6\%) & 3.9 (8.4\%) & 6.5 (14.0\%) & 41.0 & 17.8 \\
 & BOIN-Comb & 19.2 & 0.247 & 34.0 & 17.8 (66.1\%) & 3.6 (11.6\%) & 6.9 (22.4\%) & 28.4 & 16.9 \\
\addlinespace[1pt]
14 & BF-BOIN-Comb & 14.4 & 0.274 & 29.1 & 28.4 (75.0\%) & 3.0 (6.8\%) & 8.6 (18.2\%) & 39.9 & 17.2 \\
 & BOIN-Comb & 13.7 & 0.216 & 40.2 & 16.5 (63.3\%) & 2.3 (7.7\%) & 9.3 (29.0\%) & 28.1 & 16.7 \\
\addlinespace[1pt]
15 & BF-BOIN-Comb & 35.1 & 0.396 & 16.2 & 22.7 (68.4\%) & 8.0 (18.8\%) & 5.8 (12.8\%) & 36.5 & 16.3 \\
 & BOIN-Comb & 37.0 & 0.392 & 24.6 & 13.4 (55.8\%) & 7.0 (22.6\%) & 6.9 (21.6\%) & 27.3 & 16.1 \\
\bottomrule
\end{tabular}%
}
\par\smallskip
\begin{minipage}{\linewidth}
\scriptsize Counts and percentages are averaged across trials. AI is calculated among trials where a dose is selected (i.e. trials ending due to excessive toxicity are excluded); Percent correct selection includes all trials, with no selection counted as a failure. Below and above MTD refer to dose combinations with true toxicity probabilities strictly below the smallest and strictly above the largest true MTD toxicity probability, respectively. Average sample size includes escalation and backfill patients; the escalation limit does not cap total BF-BOIN-Comb enrollment.
\end{minipage}
\end{table}
\end{landscape}
\clearpage
\begin{table}[p]
\centering
\caption{Scenario 16 (overly toxic scenario) results for BF-BOIN-Comb (backfill-only conflicts) and BOIN-Comb.}
\label{tab:scenario16-nstop9}
\small\setlength{\tabcolsep}{5pt}\renewcommand{\arraystretch}{1.15}
\resizebox{\linewidth}{!}{%
\begin{tabular}{@{}rlrrr@{}}
\toprule
\shortstack{Maximum\\number of\\cohorts} & Method & \shortstack{No MTD\\selected\\(\%)} & \shortstack{Mean\\sample size\\($n$)} & \shortstack{Mean trial\\duration\\(months)}\\
\midrule
10 & BF-BOIN-Comb & 54.14 & 9.47 & 4.81 \\
 & BOIN-Comb & 52.48 & 9.42 & 5.16 \\
\addlinespace[3pt]
15 & BF-BOIN-Comb & 54.10 & 9.48 & 4.82 \\
 & BOIN-Comb & 52.50 & 9.45 & 5.18 \\
\addlinespace[3pt]
20 & BF-BOIN-Comb & 54.14 & 9.48 & 4.82 \\
 & BOIN-Comb & 52.52 & 9.45 & 5.17 \\
\bottomrule
\end{tabular}%
}
\par\smallskip
\begin{minipage}{\linewidth}
\footnotesize Maximum cohorts refer to escalation cohorts, with three patients per cohort. Means are calculated across all simulated trials. Sample size includes escalation and backfill patients.
\end{minipage}
\end{table}

%% file: BFBOINComb_compare.tex
% Separate appendix section; input after the existing appendix tables.
% Requires booktabs, graphicx, pdflscape, and citation key cheung2011dose.
\subsection{Sensitivity Analysis: Adjustment Method}
\label{app:bf-conflict-comparison}
These tables compare the two adjustment methods discussed in subsection \ref{sub:sen_ana}, comparing the two different conflict resolution methods. 
\begin{landscape}
\pagestyle{empty}
\begin{table}[p]
\centering
\caption{Operating characteristics of BF-BOIN-Comb of all-lower-dose conflict resolution vs proposed (backfill-only) method with 10 cohorts (maximum escalation enrollment 30 patients), $n_{\mathrm{stop}}=9$, and 5,000 simulated trials per scenario and design. AI denotes the accuracy index\cite{cheung2011dose}.}
\label{tab:bf-conflicts-oc-30}
\scriptsize\setlength{\tabcolsep}{4pt}\renewcommand{\arraystretch}{1.0}
\resizebox{0.92\linewidth}{!}{%
\begin{tabular}{@{}rlrrrrrrrr@{}}
\toprule
\shortstack{Scenario} & Method & \shortstack{Correct\\selection\\(\%)} & \shortstack{AI} & \shortstack{Overdosing\\selection\\(\%)} & \shortstack{Patients\\below MTD\\$n$ (\%)} & \shortstack{Patients\\at MTD(s)\\$n$ (\%)} & \shortstack{Patients\\above MTD\\$n$ (\%)} & \shortstack{Average\\sample size\\($n$)} & \shortstack{Average\\duration\\(months)}\\
\midrule
1 & BF-BOIN-Comb (all lower) & 52.3 & 0.510 & 11.1 & 29.9 (71.6\%) & 9.2 (20.7\%) & 3.6 (7.7\%) & 42.7 & 16.6 \\
 & BF-BOIN-Comb (proposed) & 52.9 & 0.516 & 11.4 & 29.8 (70.8\%) & 9.4 (21.2\%) & 3.8 (8.0\%) & 43.0 & 16.6 \\
\addlinespace[1pt]
2 & BF-BOIN-Comb (all lower) & 60.6 & 0.749 & 19.5 & 10.3 (42.9\%) & 11.7 (39.3\%) & 6.1 (17.8\%) & 28.1 & 13.1 \\
 & BF-BOIN-Comb (proposed) & 61.3 & 0.752 & 20.3 & 10.1 (41.5\%) & 11.7 (39.5\%) & 6.5 (18.9\%) & 28.4 & 13.2 \\
\addlinespace[1pt]
3 & BF-BOIN-Comb (all lower) & 48.2 & 0.467 & 12.2 & 31.8 (75.2\%) & 7.9 (18.1\%) & 3.1 (6.7\%) & 42.7 & 17.4 \\
 & BF-BOIN-Comb (proposed) & 50.1 & 0.487 & 11.1 & 31.6 (74.9\%) & 8.1 (18.5\%) & 3.0 (6.6\%) & 42.7 & 17.4 \\
\addlinespace[1pt]
4 & BF-BOIN-Comb (all lower) & 59.9 & 0.854 & 27.5 & 0.0 (0.0\%) & 9.6 (69.5\%) & 7.5 (30.5\%) & 17.0 & 9.2 \\
 & BF-BOIN-Comb (proposed) & 59.2 & 0.850 & 28.2 & 0.0 (0.0\%) & 9.5 (69.1\%) & 7.6 (30.9\%) & 17.1 & 9.3 \\
\addlinespace[1pt]
5 & BF-BOIN-Comb (all lower) & 43.4 & 0.535 & 0.0 & 39.1 (89.7\%) & 4.5 (10.3\%) & 0.0 (0.0\%) & 43.6 & 18.2 \\
 & BF-BOIN-Comb (proposed) & 45.2 & 0.551 & 0.0 & 38.9 (89.5\%) & 4.6 (10.5\%) & 0.0 (0.0\%) & 43.6 & 18.2 \\
\addlinespace[1pt]
6 & BF-BOIN-Comb (all lower) & 39.5 & 0.403 & 15.0 & 33.9 (76.9\%) & 6.5 (14.3\%) & 4.2 (8.8\%) & 44.6 & 17.1 \\
 & BF-BOIN-Comb (proposed) & 40.5 & 0.413 & 14.5 & 33.7 (76.4\%) & 6.6 (14.6\%) & 4.3 (9.0\%) & 44.6 & 17.1 \\
\addlinespace[1pt]
7 & BF-BOIN-Comb (all lower) & 52.8 & 0.582 & 15.7 & 23.2 (62.5\%) & 8.9 (24.2\%) & 5.4 (13.4\%) & 37.5 & 15.9 \\
 & BF-BOIN-Comb (proposed) & 54.7 & 0.602 & 15.5 & 22.8 (61.4\%) & 9.0 (24.8\%) & 5.6 (13.8\%) & 37.4 & 15.8 \\
\addlinespace[1pt]
8 & BF-BOIN-Comb (all lower) & 30.0 & 0.511 & 23.0 & 29.0 (73.3\%) & 4.6 (11.3\%) & 6.4 (15.4\%) & 40.0 & 16.7 \\
 & BF-BOIN-Comb (proposed) & 30.0 & 0.510 & 23.4 & 28.9 (72.5\%) & 4.8 (11.7\%) & 6.6 (15.9\%) & 40.4 & 16.8 \\
\addlinespace[1pt]
9 & BF-BOIN-Comb (all lower) & 32.7 & 0.487 & 19.5 & 31.3 (73.3\%) & 5.1 (13.4\%) & 5.8 (13.3\%) & 42.2 & 17.3 \\
 & BF-BOIN-Comb (proposed) & 32.0 & 0.479 & 19.4 & 31.2 (72.8\%) & 5.1 (13.4\%) & 5.9 (13.8\%) & 42.3 & 17.3 \\
\addlinespace[1pt]
10 & BF-BOIN-Comb (all lower) & 20.4 & 0.576 & 42.8 & 23.9 (63.2\%) & 2.6 (6.0\%) & 10.9 (30.8\%) & 37.5 & 16.1 \\
 & BF-BOIN-Comb (proposed) & 22.2 & 0.584 & 42.4 & 24.1 (62.5\%) & 3.0 (6.7\%) & 11.1 (30.8\%) & 38.1 & 16.1 \\
\addlinespace[1pt]
11 & BF-BOIN-Comb (all lower) & 89.5 & 0.428 & 0.0 & 25.6 (74.0\%) & 8.4 (26.0\%) & 0.0 (0.0\%) & 34.1 & 15.1 \\
 & BF-BOIN-Comb (proposed) & 89.5 & 0.427 & 0.0 & 25.5 (73.9\%) & 8.5 (26.1\%) & 0.0 (0.0\%) & 34.0 & 15.1 \\
\addlinespace[1pt]
12 & BF-BOIN-Comb (all lower) & 53.3 & 0.398 & 0.0 & 26.2 (74.1\%) & 9.5 (25.9\%) & 0.0 (0.0\%) & 35.7 & 15.9 \\
 & BF-BOIN-Comb (proposed) & 54.7 & 0.413 & 0.0 & 26.1 (73.4\%) & 9.8 (26.6\%) & 0.0 (0.0\%) & 35.9 & 16.0 \\
\addlinespace[1pt]
13 & BF-BOIN-Comb (all lower) & 15.1 & 0.210 & 21.9 & 27.1 (79.6\%) & 3.1 (8.0\%) & 4.9 (12.4\%) & 35.1 & 15.4 \\
 & BF-BOIN-Comb (proposed) & 16.1 & 0.228 & 22.6 & 27.2 (78.1\%) & 3.3 (8.5\%) & 5.3 (13.4\%) & 35.9 & 15.5 \\
\addlinespace[1pt]
14 & BF-BOIN-Comb (all lower) & 15.1 & 0.266 & 26.8 & 25.5 (77.1\%) & 2.4 (6.5\%) & 6.5 (16.4\%) & 34.3 & 14.7 \\
 & BF-BOIN-Comb (proposed) & 14.5 & 0.260 & 27.4 & 25.1 (76.4\%) & 2.4 (6.6\%) & 6.7 (16.9\%) & 34.2 & 14.6 \\
\addlinespace[1pt]
15 & BF-BOIN-Comb (all lower) & 33.2 & 0.386 & 17.1 & 20.7 (70.1\%) & 6.5 (17.6\%) & 4.7 (12.3\%) & 31.9 & 14.0 \\
 & BF-BOIN-Comb (proposed) & 34.3 & 0.387 & 16.5 & 20.2 (69.3\%) & 6.6 (18.0\%) & 4.9 (12.7\%) & 31.7 & 13.9 \\
\bottomrule
\end{tabular}%
}
\par\smallskip
\begin{minipage}{\linewidth}
\scriptsize Counts and percentages are averaged across trials. AI is calculated among trials where a dose is selected (i.e. trials ending due to excessive toxicity are excluded); Percent correct selection includes all trials, with no selection counted as a failure. Below and above MTD refer to dose combinations with true toxicity probabilities strictly below the smallest and strictly above the largest true MTD toxicity probability, respectively. Average sample size includes escalation and backfill patients; the escalation limit does not cap total BF-BOIN-Comb enrollment.
\end{minipage}
\end{table}
\end{landscape}
\clearpage
\begin{landscape}
\pagestyle{empty}
\begin{table}[p]
\centering
\caption{Operating characteristics of BF-BOIN-Comb of all-lower-dose conflict resolution vs proposed (backfill-only) method with 15 cohorts (maximum escalation enrollment 45 patients), $n_{\mathrm{stop}}=9$, and 5,000 simulated trials per scenario and design. AI denotes the accuracy index\cite{cheung2011dose}.}
\label{tab:bf-conflicts-oc-45}
\scriptsize\setlength{\tabcolsep}{4pt}\renewcommand{\arraystretch}{1.0}
\resizebox{0.92\linewidth}{!}{%
\begin{tabular}{@{}rlrrrrrrrr@{}}
\toprule
\shortstack{Scenario} & Method & \shortstack{Correct\\selection\\(\%)} & \shortstack{AI} & \shortstack{Overdosing\\selection\\(\%)} & \shortstack{Patients\\below MTD\\$n$ (\%)} & \shortstack{Patients\\at MTD(s)\\$n$ (\%)} & \shortstack{Patients\\above MTD\\$n$ (\%)} & \shortstack{Average\\sample size\\($n$)} & \shortstack{Average\\duration\\(months)}\\
\midrule
1 & BF-BOIN-Comb (all lower) & 55.4 & 0.540 & 10.8 & 34.3 (70.0\%) & 11.6 (21.8\%) & 4.8 (8.3\%) & 50.7 & 19.5 \\
 & BF-BOIN-Comb (proposed) & 54.8 & 0.535 & 11.0 & 33.8 (69.5\%) & 11.6 (21.8\%) & 5.1 (8.7\%) & 50.5 & 19.4 \\
\addlinespace[1pt]
2 & BF-BOIN-Comb (all lower) & 62.3 & 0.758 & 20.0 & 11.0 (41.5\%) & 13.2 (39.7\%) & 7.2 (18.8\%) & 31.4 & 14.8 \\
 & BF-BOIN-Comb (proposed) & 61.9 & 0.756 & 20.2 & 10.8 (40.8\%) & 13.1 (39.9\%) & 7.5 (19.4\%) & 31.4 & 14.8 \\
\addlinespace[1pt]
3 & BF-BOIN-Comb (all lower) & 54.8 & 0.535 & 10.1 & 36.7 (72.6\%) & 10.7 (19.9\%) & 4.3 (7.5\%) & 51.8 & 20.7 \\
 & BF-BOIN-Comb (proposed) & 55.1 & 0.540 & 10.2 & 36.5 (71.6\%) & 11.0 (20.4\%) & 4.6 (8.0\%) & 52.2 & 20.9 \\
\addlinespace[1pt]
4 & BF-BOIN-Comb (all lower) & 61.3 & 0.859 & 26.8 & 0.0 (0.0\%) & 10.0 (69.5\%) & 8.1 (30.5\%) & 18.1 & 9.9 \\
 & BF-BOIN-Comb (proposed) & 61.1 & 0.861 & 26.3 & 0.0 (0.0\%) & 9.9 (69.6\%) & 8.2 (30.4\%) & 18.1 & 9.9 \\
\addlinespace[1pt]
5 & BF-BOIN-Comb (all lower) & 53.0 & 0.613 & 0.0 & 45.0 (87.2\%) & 6.6 (12.8\%) & 0.0 (0.0\%) & 51.6 & 21.0 \\
 & BF-BOIN-Comb (proposed) & 53.7 & 0.620 & 0.0 & 44.7 (86.9\%) & 6.8 (13.1\%) & 0.0 (0.0\%) & 51.4 & 21.0 \\
\addlinespace[1pt]
6 & BF-BOIN-Comb (all lower) & 43.3 & 0.447 & 15.0 & 39.7 (74.8\%) & 8.5 (15.5\%) & 5.8 (9.8\%) & 54.1 & 20.5 \\
 & BF-BOIN-Comb (proposed) & 43.1 & 0.446 & 14.3 & 39.3 (74.1\%) & 8.7 (15.8\%) & 6.0 (10.1\%) & 54.0 & 20.4 \\
\addlinespace[1pt]
7 & BF-BOIN-Comb (all lower) & 60.5 & 0.654 & 13.5 & 26.4 (60.9\%) & 11.2 (26.1\%) & 6.3 (13.0\%) & 43.8 & 18.7 \\
 & BF-BOIN-Comb (proposed) & 60.2 & 0.650 & 13.8 & 25.9 (59.6\%) & 11.3 (26.2\%) & 7.0 (14.2\%) & 44.1 & 18.7 \\
\addlinespace[1pt]
8 & BF-BOIN-Comb (all lower) & 31.5 & 0.523 & 23.0 & 33.7 (71.5\%) & 6.1 (12.2\%) & 8.3 (16.3\%) & 48.1 & 20.0 \\
 & BF-BOIN-Comb (proposed) & 33.1 & 0.536 & 20.6 & 33.3 (71.0\%) & 6.3 (12.8\%) & 8.3 (16.3\%) & 48.0 & 19.9 \\
\addlinespace[1pt]
9 & BF-BOIN-Comb (all lower) & 37.6 & 0.543 & 18.2 & 37.4 (71.7\%) & 6.6 (14.4\%) & 7.4 (13.9\%) & 51.5 & 20.8 \\
 & BF-BOIN-Comb (proposed) & 35.0 & 0.525 & 17.8 & 37.5 (71.7\%) & 6.3 (14.0\%) & 7.6 (14.2\%) & 51.5 & 20.7 \\
\addlinespace[1pt]
10 & BF-BOIN-Comb (all lower) & 28.5 & 0.623 & 41.6 & 28.8 (61.1\%) & 4.6 (8.5\%) & 13.5 (30.4\%) & 46.9 & 20.5 \\
 & BF-BOIN-Comb (proposed) & 31.1 & 0.636 & 40.6 & 28.2 (60.1\%) & 5.2 (9.5\%) & 13.6 (30.4\%) & 47.0 & 20.1 \\
\addlinespace[1pt]
11 & BF-BOIN-Comb (all lower) & 91.2 & 0.434 & 0.0 & 26.2 (73.8\%) & 8.7 (26.2\%) & 0.0 (0.0\%) & 34.9 & 15.4 \\
 & BF-BOIN-Comb (proposed) & 91.1 & 0.434 & 0.0 & 26.1 (73.7\%) & 8.7 (26.3\%) & 0.0 (0.0\%) & 34.8 & 15.4 \\
\addlinespace[1pt]
12 & BF-BOIN-Comb (all lower) & 57.3 & 0.425 & 0.0 & 28.8 (73.2\%) & 11.0 (26.8\%) & 0.0 (0.0\%) & 39.8 & 17.7 \\
 & BF-BOIN-Comb (proposed) & 56.3 & 0.419 & 0.0 & 28.5 (72.8\%) & 11.2 (27.2\%) & 0.0 (0.0\%) & 39.7 & 17.6 \\
\addlinespace[1pt]
13 & BF-BOIN-Comb (all lower) & 15.2 & 0.221 & 21.8 & 30.7 (79.2\%) & 3.6 (8.1\%) & 5.8 (12.8\%) & 40.1 & 17.6 \\
 & BF-BOIN-Comb (proposed) & 15.3 & 0.222 & 22.4 & 30.2 (77.6\%) & 3.9 (8.7\%) & 6.3 (13.7\%) & 40.3 & 17.4 \\
\addlinespace[1pt]
14 & BF-BOIN-Comb (all lower) & 14.9 & 0.265 & 27.2 & 27.9 (76.5\%) & 2.9 (6.8\%) & 7.5 (16.7\%) & 38.3 & 16.6 \\
 & BF-BOIN-Comb (proposed) & 14.6 & 0.268 & 28.4 & 27.5 (75.6\%) & 2.9 (6.9\%) & 8.0 (17.6\%) & 38.4 & 16.5 \\
\addlinespace[1pt]
15 & BF-BOIN-Comb (all lower) & 34.3 & 0.390 & 17.0 & 22.3 (69.4\%) & 7.5 (18.2\%) & 5.4 (12.4\%) & 35.2 & 15.7 \\
 & BF-BOIN-Comb (proposed) & 35.6 & 0.396 & 16.8 & 22.1 (68.3\%) & 7.8 (18.9\%) & 5.6 (12.8\%) & 35.5 & 15.7 \\
\bottomrule
\end{tabular}%
}
\par\smallskip
\begin{minipage}{\linewidth}
\scriptsize Counts and percentages are averaged across trials.
AI is calculated among trials where a dose is selected (i.e. trials ending due to excessive toxicity are excluded); Percent correct selection includes all trials, with no selection counted as a failure. Below and above MTD refer to dose combinations with true toxicity probabilities strictly below the smallest and strictly above the largest true MTD toxicity probability, respectively. Average sample size includes escalation and backfill patients; the escalation limit does not cap total BF-BOIN-Comb enrollment.
\end{minipage}
\end{table}
\end{landscape}
\clearpage
\begin{landscape}
\pagestyle{empty}
\begin{table}[p]
\centering
\caption{Operating characteristics of BF-BOIN-Comb of all-lower-dose conflict resolution vs proposed (backfill-only) method with 20 cohorts (maximum escalation enrollment 60 patients), $n_{\mathrm{stop}}=9$, and 5,000 simulated trials per scenario and design. AI denotes the accuracy index\cite{cheung2011dose}.}
\label{tab:bf-conflicts-oc-60}
\scriptsize\setlength{\tabcolsep}{4pt}\renewcommand{\arraystretch}{1.0}
\resizebox{0.92\linewidth}{!}{%
\begin{tabular}{@{}rlrrrrrrrr@{}}
\toprule
\shortstack{Scenario} & Method & \shortstack{Correct\\selection\\(\%)} & \shortstack{AI} & \shortstack{Overdosing\\selection\\(\%)} & \shortstack{Patients\\below MTD\\$n$ (\%)} & \shortstack{Patients\\at MTD(s)\\$n$ (\%)} & \shortstack{Patients\\above MTD\\$n$ (\%)} & \shortstack{Average\\sample size\\($n$)} & \shortstack{Average\\duration\\(months)}\\
\midrule
1 & BF-BOIN-Comb (all lower) & 55.5 & 0.544 & 11.5 & 35.7 (69.6\%) & 12.4 (22.0\%) & 5.2 (8.4\%) & 53.3 & 20.6 \\
 & BF-BOIN-Comb (proposed) & 56.2 & 0.550 & 10.7 & 34.9 (68.8\%) & 12.6 (22.3\%) & 5.5 (8.9\%) & 52.9 & 20.4 \\
\addlinespace[1pt]
2 & BF-BOIN-Comb (all lower) & 62.3 & 0.759 & 19.8 & 11.2 (41.3\%) & 13.7 (39.9\%) & 7.5 (18.8\%) & 32.4 & 15.5 \\
 & BF-BOIN-Comb (proposed) & 62.9 & 0.762 & 19.8 & 11.1 (40.7\%) & 13.7 (40.0\%) & 7.8 (19.3\%) & 32.6 & 15.5 \\
\addlinespace[1pt]
3 & BF-BOIN-Comb (all lower) & 55.6 & 0.547 & 10.5 & 38.5 (72.2\%) & 11.6 (20.2\%) & 4.6 (7.5\%) & 54.7 & 21.8 \\
 & BF-BOIN-Comb (proposed) & 55.5 & 0.545 & 10.2 & 38.0 (71.3\%) & 12.0 (20.7\%) & 4.9 (7.9\%) & 54.9 & 21.9 \\
\addlinespace[1pt]
4 & BF-BOIN-Comb (all lower) & 60.7 & 0.857 & 27.0 & 0.0 (0.0\%) & 10.2 (69.2\%) & 8.5 (30.8\%) & 18.6 & 10.1 \\
 & BF-BOIN-Comb (proposed) & 61.0 & 0.860 & 26.4 & 0.0 (0.0\%) & 10.0 (69.3\%) & 8.4 (30.7\%) & 18.5 & 10.0 \\
\addlinespace[1pt]
5 & BF-BOIN-Comb (all lower) & 54.3 & 0.626 & 0.0 & 46.2 (86.9\%) & 6.9 (13.1\%) & 0.0 (0.0\%) & 53.1 & 21.6 \\
 & BF-BOIN-Comb (proposed) & 55.2 & 0.634 & 0.0 & 45.8 (86.5\%) & 7.2 (13.5\%) & 0.0 (0.0\%) & 52.9 & 21.6 \\
\addlinespace[1pt]
6 & BF-BOIN-Comb (all lower) & 43.6 & 0.450 & 13.6 & 42.2 (74.9\%) & 9.3 (15.7\%) & 6.0 (9.4\%) & 57.6 & 21.8 \\
 & BF-BOIN-Comb (proposed) & 43.1 & 0.443 & 15.2 & 41.2 (73.9\%) & 9.3 (15.9\%) & 6.6 (10.3\%) & 57.1 & 21.6 \\
\addlinespace[1pt]
7 & BF-BOIN-Comb (all lower) & 62.3 & 0.669 & 13.4 & 27.4 (59.9\%) & 12.1 (26.6\%) & 6.9 (13.4\%) & 46.4 & 19.9 \\
 & BF-BOIN-Comb (proposed) & 63.4 & 0.679 & 13.0 & 26.7 (59.2\%) & 12.1 (26.8\%) & 7.1 (14.0\%) & 45.9 & 19.6 \\
\addlinespace[1pt]
8 & BF-BOIN-Comb (all lower) & 34.1 & 0.543 & 20.6 & 35.2 (70.9\%) & 7.1 (13.2\%) & 8.7 (15.9\%) & 51.0 & 21.4 \\
 & BF-BOIN-Comb (proposed) & 33.8 & 0.543 & 21.2 & 34.8 (70.5\%) & 7.0 (13.0\%) & 9.0 (16.4\%) & 50.8 & 21.2 \\
\addlinespace[1pt]
9 & BF-BOIN-Comb (all lower) & 40.2 & 0.564 & 16.4 & 39.4 (71.3\%) & 7.5 (15.3\%) & 7.6 (13.4\%) & 54.5 & 22.1 \\
 & BF-BOIN-Comb (proposed) & 39.4 & 0.557 & 16.6 & 38.5 (70.8\%) & 7.2 (15.1\%) & 7.9 (14.1\%) & 53.6 & 21.7 \\
\addlinespace[1pt]
10 & BF-BOIN-Comb (all lower) & 32.5 & 0.648 & 40.2 & 30.4 (60.3\%) & 5.8 (9.9\%) & 14.4 (29.9\%) & 50.6 & 22.4 \\
 & BF-BOIN-Comb (proposed) & 33.5 & 0.653 & 39.3 & 29.8 (59.6\%) & 6.1 (10.3\%) & 14.4 (30.1\%) & 50.3 & 21.8 \\
\addlinespace[1pt]
11 & BF-BOIN-Comb (all lower) & 91.2 & 0.434 & 0.0 & 26.3 (73.7\%) & 8.7 (26.3\%) & 0.0 (0.0\%) & 35.0 & 15.5 \\
 & BF-BOIN-Comb (proposed) & 91.4 & 0.435 & 0.0 & 26.1 (73.6\%) & 8.8 (26.4\%) & 0.0 (0.0\%) & 34.9 & 15.4 \\
\addlinespace[1pt]
12 & BF-BOIN-Comb (all lower) & 57.3 & 0.428 & 0.0 & 29.4 (73.1\%) & 11.3 (26.9\%) & 0.0 (0.0\%) & 40.7 & 18.1 \\
 & BF-BOIN-Comb (proposed) & 57.7 & 0.432 & 0.0 & 29.3 (72.4\%) & 11.6 (27.6\%) & 0.0 (0.0\%) & 40.9 & 18.1 \\
\addlinespace[1pt]
13 & BF-BOIN-Comb (all lower) & 15.9 & 0.227 & 23.3 & 31.3 (78.5\%) & 3.8 (8.4\%) & 6.1 (13.1\%) & 41.2 & 18.1 \\
 & BF-BOIN-Comb (proposed) & 15.4 & 0.228 & 23.6 & 30.7 (77.6\%) & 3.9 (8.4\%) & 6.5 (14.0\%) & 41.0 & 17.8 \\
\addlinespace[1pt]
14 & BF-BOIN-Comb (all lower) & 14.4 & 0.271 & 27.6 & 28.8 (76.3\%) & 2.9 (6.6\%) & 8.0 (17.0\%) & 39.8 & 17.4 \\
 & BF-BOIN-Comb (proposed) & 14.4 & 0.274 & 29.1 & 28.4 (75.0\%) & 3.0 (6.8\%) & 8.6 (18.2\%) & 39.9 & 17.2 \\
\addlinespace[1pt]
15 & BF-BOIN-Comb (all lower) & 36.3 & 0.402 & 15.9 & 23.0 (69.6\%) & 7.9 (18.5\%) & 5.4 (11.9\%) & 36.2 & 16.2 \\
 & BF-BOIN-Comb (proposed) & 35.1 & 0.396 & 16.2 & 22.7 (68.4\%) & 8.0 (18.8\%) & 5.8 (12.8\%) & 36.5 & 16.3 \\
\bottomrule
\end{tabular}%
}
\par\smallskip
\begin{minipage}{\linewidth}
\scriptsize Counts and percentages are averaged across trials. AI is calculated among trials where a dose is selected (i.e. trials ending due to excessive toxicity are excluded); Percent correct selection includes all trials, with no selection counted as a failure. Below and above MTD refer to dose combinations with true toxicity probabilities strictly below the smallest and strictly above the largest true MTD toxicity probability, respectively. Average sample size includes escalation and backfill patients; the escalation limit does not cap total BF-BOIN-Comb enrollment.
\end{minipage}
\end{table}
\end{landscape}
\clearpage
\begin{table}[p]
\centering
\caption{Scenario 16 (overly toxic scenario) results for BF-BOIN-Comb with all-lower-dose conflict resolution vs. proposed method.}
\label{tab:bf-conflicts-scenario16-nstop9}
\small\setlength{\tabcolsep}{5pt}\renewcommand{\arraystretch}{1.15}
\resizebox{\linewidth}{!}{%
\begin{tabular}{@{}rlrrr@{}}
\toprule
\shortstack{Maximum\\number of\\cohorts} & Method & \shortstack{No MTD\\selected\\(\%)} & \shortstack{Mean\\sample size\\($n$)} & \shortstack{Mean trial\\duration\\(months)}\\
\midrule
10 & BF-BOIN-Comb (all lower) & 54.22 & 9.45 & 4.80 \\
 & BF-BOIN-Comb (proposed) & 54.14 & 9.47 & 4.81 \\
\addlinespace[3pt]
15 & BF-BOIN-Comb (all lower) & 54.18 & 9.47 & 4.81 \\
 & BF-BOIN-Comb (proposed) & 54.10 & 9.48 & 4.82 \\
\addlinespace[3pt]
20 & BF-BOIN-Comb (all lower) & 54.18 & 9.47 & 4.81 \\
 & BF-BOIN-Comb (proposed) & 54.14 & 9.48 & 4.82 \\
\bottomrule
\end{tabular}%
}
\par\smallskip
\begin{minipage}{\linewidth}
\footnotesize Maximum cohorts refer to escalation cohorts, with three patients per cohort. Means are calculated across all simulated trials. Sample size includes escalation and backfill patients.
\end{minipage}
\end{table}